\documentclass[final,5p,times,twocolumn,authoryear]{elsarticle}

\usepackage{amssymb}
\usepackage{amsmath}
\usepackage{bm}
\usepackage[caption=false,font=normalsize,labelfont=sf,textfont=sf]{subfig}
\usepackage[switch]{lineno}
\usepackage{amsmath}
\usepackage{makecell}
\usepackage[hidelinks]{hyperref}
\usepackage{url}
\hypersetup{breaklinks=true}

\newtheorem{definition}{Definition}

\newcommand{\revision}[1]{{\color{black} #1}}

\usepackage[some]{background}
\SetBgScale{1}
\SetBgContents{\parbox{20cm}{%
  \Huge Draft:  \today\\[18cm]\rotatebox{180}{\Huge Draft:  \today}}}
\SetBgColor{red}
\SetBgAngle{270}
\SetBgOpacity{0.2}

\begin{document}

\begin{frontmatter}



\author{Federico Zocco\corref{cor1}}
\ead{federico.zocco.fz@gmail.com}
\author{Andrea Corti}
\ead{andrea.corti@unisi.it}
\author{Monica Malvezzi}
\ead{monica.malvezzi@unisi.it}

\cortext[cor1]{Corresponding author}
\affiliation{organization={Department of Information Engineering and Mathematics, University of Siena},
             addressline={Via Roma 56},
             city={Siena},
             postcode={53100},
             country={Italy}}

\title{CiRL: Open-Source Environments for Reinforcement Learning in Circular Economy and Net Zero}

\begin{abstract}
The demand of finite raw materials  will keep increasing as they fuel modern society. Simultaneously, solutions for stopping carbon emissions in the short term are not available, thus making the net zero target extremely challenging to achieve at scale. The circular economy (CE) paradigm is gaining attention as a solution to address climate change and the uncertainties of supplies of critical materials. Hence, in this paper, we introduce CiRL, a deep reinforcement learning (DRL) library of environments focused on the circularity \revision{control} of both solid and fluid materials. The integration of DRL into the design of material circularity is possible thanks to the formalism of thermodynamical material networks, which is underpinned by compartmental dynamical thermodynamics. Along with the focus on circularity, this library has three more features: the new CE-oriented environments are in the state-space form, which is typically used in dynamical systems analysis and control design; it is based on a state-of-the-art Python library of DRL algorithms, namely, Stable-Baselines3; and it is developed in Google Colaboratory to be accessible to researchers from different disciplines and backgrounds as is often the case for circular economy researchers and engineers. \revision{CiRL is intended to be a tool to generate AI-driven actions for optimizing the circularity of supply-recovery chains and to be combined with human-driven decisions derived from material flow analysis (MFA) studies.} CiRL is publicly available\footnotemark.
\end{abstract}



\begin{keyword}
Waste sorting \sep air pollution control \sep nature-based solutions \sep circular robotics \sep circular intelligence.
\end{keyword}

\end{frontmatter}


\footnotetext[1]{\url{https://github.com/ciroresearch/CiRL}}

\section{Introduction}
While modern societies provide customized and effective services and products to individuals, they rely on strategic materials whose supply is not guaranteed in the long-term (say, in fifty years) because of the finiteness of the natural reserves, the prohibitive costs for further extraction, or the unreliability of the countries from which they are imported \citep{nygaard2023geopolitical,CRM-EU}. Another open issue of modern societies is the generation of waste that ends-up in water bodies \citep{suryawan2024reshaping} and urban areas \citep{pietrelli2025characterization,ballatore2022city}. In addition, carbon dioxide emissions resulting from combustion in energy, heating, and aerospace systems continue to increase since today there is no immediate solution to stop them \citep{NASA-ClimateChange}. Hence, the net-zero target requested by several governments to prevent a climate event horizon is extremely challenging to reach \citep{NZ-DefUniOx,EU-NZ,UK-NZ,US-NZ}. 
The coexistence of these problems motivates the rise of the paradigm of circular economy (CE), in which solid and fluid materials are kept in use for as long as possible through reusing, repairing, refurbishing, remanufacturing, and also through reducing the material demand itself \citep{EllenMacArthurFound,liaros2021circular,ossio2023circular,potting2017circular,pajunen2022circular}. 

Deep reinforcement learning (DRL) is a very promising control approach since it enhances classical control techniques by leveraging the latest advances in deep learning. Although not yet performing for highly dynamic real systems, DRL has the potential to yield very adaptive and reliable controllers if combined with the stability property developed in control theory \citep{bucsoniu2018reinforcement,han2020actor}. \revision{However, challenges remain open such as the sample inefficiency of learning algorithms \citep{zheng2024sample} and the formulation of effective reward functions to solve real-world problems \citep{feng2025deep,lanzaro2025evaluating}.} The interest for DRL is demonstrated by the proliferation of libraries and physical simulators created in the last four years \citep{fujita2021chainerrl,weng2022tianshou,seno2022d3rlpy,tarasov2023corl,serrano2023skrl,gallouedec2021panda,scheikl2023lapgym,suarez2024pufferlib,raffin2021stable,kaup2024review}.

Despite the above-mentioned benefits of a transition to a circular economy and the potential of DRL for effective decision making, the link between CE and DRL remains, to the best of our knowledge, still vague in the literature mainly because of the lack of clarity on how to measure the circularity of a system. For example, \cite{halter2025techno} found 798 circularity indicators used in the literature. In other words, it is unanswered the question: ``How the recent advances in DRL can increase circularity?''   

In this context, this paper makes the following main contributions:
\begin{itemize}
\item{It improves a quantitative dynamical physics-based measure of circularity, namely, $\lambda(\mathcal{N}; t)$, and the circularity optimization problem, that is, 
\begin{equation}\label{eq:optimizeCircularityContrib}
\mathcal{N}^* = \arg \max \,\, \lambda(\mathcal{N}; t),    
\end{equation}
where $\mathcal{N}$ is a network of thermodynamic compartments that exchange and process a target material, e.g., plastic, gold, carbon dioxide \citep{zocco2024unification}.}
\item{It provides an open-source library of DRL environments to solve (\ref{eq:optimizeCircularityContrib}), namely, \emph{CiRL}, leveraging the state-of-the-art DRL algorithm implementations of Stable-Baselines3 \citep{raffin2021stable}. In addition, CiRL is user-friendly to be accessible to the circular economy community and its environments are in the state-space form to highlight the parallelism with classical control theory, and hence, to reach the classical control community.}
\end{itemize}

The paper adopts the following notation: vectors and matrices are indicated with bold lower-case and capital letters, respectively, while the sets are indicated with calligraphic letters. 

The rest of the paper is organized as follows: Section \ref{sec:RelWork} covers the related work, then Section \ref{sec:Preliminaries} provides the preliminary definitions and concepts necessary to proceed; subsequently, Section \ref{sec:Features} details the features of CiRL, Section \ref{sec:Compartments} explains the environments implemented in CiRL so far, while Section \ref{sec:forUseAndDev} gives suggestions to users and developers of CiRL. \revision{Finally, Section \ref{sec:ApplicationsAndImpact} discusses the applications of CiRL, Section \ref{sec:Limitations} discusses the limitations and future work, and Section \ref{sec:Conclusion} gives the conclusion.}

\section{Related Work}\label{sec:RelWork}
\subsection{Libraries for Deep Reinforcement Learning}
Several DRL libraries have been proposed with different features and applications. \revision{in this section, we mention some of the libraries proposed from 2016 to 2024 following a chronological order.} One of the first DRL libraries was \emph{KerasRL} \citep{plappert2016kerasrl} proposed in 2016 and it is based on the deep learning framework Keras. One of the most popular and extensive libraries is \emph{SB3} \citep{raffin2021stable} and was proposed in 2021. Another library proposed in 2021 was \emph{ChainerRL} \citep{fujita2021chainerrl}, which was based on the deep learning framework Chainer. \revision{RL algorithms can work either off-line or on-line; the former train the agents using datasets of historical data of observations, actions, and rewards; thus, the dataset defines the environment in the off-line setting. In contrast, on-line methods require that the training data are generated while the agent is learning. CiRL follows the mechanics of the latter.} Two libraries proposed in 2022 are \emph{Tianshou} \citep{weng2022tianshou} and \emph{d3rlpy} \citep{seno2022d3rlpy}. The former supports both off-line and on-line training through a unified interface, while the latter focuses on off-line learning. Another library focused on off-line settings is \emph{CORL} \citep{tarasov2023corl}, which was proposed in 2023. \emph{CORL} considers also the offline-to-online option. A library supporting a particularly vast set of environment interfaces is \emph{skrl} because it supports, along with the popular Farama Gymnasium interface \citep{towers2024gymnasium}, the NVIDIA Isaac Gym, Isaac Orbit, and Omniverse Isaac Gym environments \citep{serrano2023skrl}. \revision{To date, Isaac Gym is no longer under development, all development and features are in IsaacSim. Isaac Orbit is open source so that contributions can be made by the community; Isaac Orbit now evolves as Isaac Lab. Our library is compatible with the Farama Gymnasium interface only.} The libraries mentioned so far do not target specific domains of applications. In contrast, \emph{FinRL} focuses on quantitative finance \citep{liu2020finrl}, \emph{Panda-Gym} focuses on the Franka Emika Panda robot \citep{gallouedec2021panda}, while \emph{LapGym} focuses on robot-assisted laparoscopic surgery \citep{scheikl2023lapgym}. A library proposed in 2024 is \emph{PufferLib}, whose aim is to facilitate the combination of the algorithms available in libraries, e.g, \emph{SB3}, and environment simulators \citep{suarez2024pufferlib}. An extensive list of libraries is available in \citep{listOfRLlibraries}, while a comparison of selected ones is provided by \citep{nouwou2023comparison}.

With respect to existing libraries, \emph{CiRL}: is largely developed through Python notebooks similarly to \emph{FinRL}, it follows the Open AI Gym/Farama Gymnasium interface similarly to \emph{SB3}, \emph{ChainerRL}, and \emph{d3rlpy}, and it relies on \emph{SB3} for the training algorithms; \emph{CiRL} is the first library aiming to optimize the circularity of materials, namely, $\lambda(\mathcal{N}; t)$ \citep{zocco2024circular}. \revision{What makes CiRL unique is that, to date, is the only library that supports decision making in circular systems designs. Furthermore, it is the only library of environments written in the state-space form; this is to stimulate analysis from and discussions with control theorists, whose control techniques differ from RL. At a high-level, CiRL helps with the European effort to develop digital solutions for a circular economy as promoted, for example, by the DICE Network+ in the United Kingdom \citep{DICENetwork+} and by the DICE Lab in central Europe \citep{DiCE-Lab}.}

\subsection{Circular Economy}
The paradigm of circular economy is gaining interest as a solution to reduce material supply uncertainties \citep{nygaard2023geopolitical} and waste generation \citep{luttenberger2020waste} by closing the loop of material flows, e.g., the output materials of a process can enter another process as secondary raw material. \emph{Circularity} extends the material life-cycle and reduces the demand of finite raw materials, especially those imported from uncertain natural reserves or unreliable areas of the world \citep{distefano2024material}. 

At the foundations of circularity are the so called ``Rs'', that indicate the practices to be adopted in order to increase circularity. Examples of these practices are the reduction of material demand, the reuse of materials and product components, the repair of parts, and recycling \citep{potting2017circular}. The ``Rs'' are usually ordered from the least to the most energy intensive and the aim is to extend the life-cycle of products and materials at their highest value possible (thus, recycling is one of the last options). The adoption of the ``Rs'' is being investigated in different sectors, e.g., in healthcare \citep{van2021circular}, in automotive \citep{prochatzki2023critical}, in aerospace \citep{dias2022possibilities}, in food \citep{liaros2021circular}, in fashion \citep{peleg2022regulation}, in electronics \citep{pajunen2022circular}, and in construction \citep{ossio2023circular}.   

While the ``Rs'' are clear practices if considered separately, a key challenge in CE is to know \emph{when}, \emph{where}, and \emph{what} ``R'' to implement in order to \emph{increase circularity of the whole material life-cycle}. At a theoretical level, answering to this question requires to design the network of processes that transport and transform the target material in order to maximize circularity. This translates into the optimization problem   
\begin{equation}\label{eq:optimizeCircularity}
\mathcal{N}^* = \arg \max \,\, \lambda(\mathcal{N}; t),    
\end{equation}
where $\mathcal{N}$ is a network of \emph{thermodynamic compartments} designed to maximize circularity $\lambda(\mathcal{N}; t)$ \citep{zocco2024circular,zocco2024unification}. The next section will explain $\mathcal{N}$ and $\lambda(\mathcal{N}; t)$ in details.

\section{Preliminaries}\label{sec:Preliminaries}
As introduced at the end of the previous section, the optimization of circularity $\lambda(\mathcal{N}; t)$ can be stated as (\ref{eq:optimizeCircularity}). We now give, in order, the definition of directed graph, of thermodynamical material network $\mathcal{N}$, and finally, of circularity $\lambda(\mathcal{N}; t)$.
\begin{definition}[\cite{bondy1976graph}]\label{def:Digraph}
A directed graph $D$ or \emph{digraph} is a graph identified by a set of $n_\text{v}$ \emph{nodes} $\{v_1, v_2, \dots, v_{n_\text{v}}\}$ and a set of $n_\text{a}$ \emph{arcs} $\{a_1, a_2, \dots, a_{n_\text{a}}\}$ that connect the nodes. A digraph $D$ in which each node or arc is associated with a \emph{weight} is a \emph{weighted digraph}. 
\end{definition}
\begin{definition}[\cite{zocco2023thermodynamical}]\label{def:TMN}
A \emph{thermodynamical material network} (TMN) is a set $\mathcal{N}$ of connected thermodynamic compartments, that is, 
\begin{equation}\label{def:TMNset}
\begin{gathered}
\mathcal{N} = \left\{c^1_{1,1}, \dots, c^{k_\text{v}}_{k_\text{v},k_\text{v}}, \dots, c^{n_\text{v}}_{n_\text{v},n_\text{v}}, \right. \\ 
\left. c^{n_\text{v}+1}_{i_{n_\text{v}+1},j_{n_\text{v}+1}}, \dots, c^{n_\text{v}+k_\text{a}}_{i_{n_\text{v}+k_\text{a}},j_{n_\text{v}+k_\text{a}}}, \dots, c^{n_\text{c}}_{i_{n_\text{c}},j_{n_\text{c}}}\right\}, 
\end{gathered}
\end{equation}
which transport, store, use, and transform a target material, namely, $\beta$. Each compartment is indicated by a \emph{control surface} \citep{moran2010fundamentals} and is modeled using \emph{dynamical systems} derived from a mass balance and/or at least one of the laws of \emph{thermodynamics} \citep{haddad2019dynamical}.
\end{definition}

Specifically, $\mathcal{N} = \mathcal{R} \cup \mathcal{T}$, where $\mathcal{R} \subseteq \mathcal{N}$ is the subset of compartments $c^k_{i,j}$ that \emph{store}, \emph{transform}, or \emph{use} the target material, while $\mathcal{T} \subset \mathcal{N}$ is the subset of compartments $c^k_{i,j}$ that \emph{move} the target material between the compartments belonging to $\mathcal{R} \subseteq \mathcal{N}$. A net $\mathcal{N}$ is associated with its \emph{compartmental diagraph} $M(\mathcal{N})$, which is a digraph whose nodes are the compartments $c^k_{i,j} \in \mathcal{R}$ and whose arcs are the compartments $c^k_{i,j} \in \mathcal{T}$. For node-compartments $c^k_{i,j} \in \mathcal{R}$ it holds that $i = j = k$, whereas for arc-compartments $c^k_{i,j} \in \mathcal{T}$ it holds that $i \neq j$ because an arc moves the material from the node-compartment $c^i_{i,i}$ to the node-compartment $c^j_{j,j}$. The orientation of an arc is given by the direction of the material flow. The superscript $k$ is the identifier of each compartment. The superscripts $k_\text{v}$ and $k_\text{a}$ in (\ref{def:TMNset}) are the $k$-th node and the $k$-th arc, respectively, while $n_\text{c}$ and $n_\text{v}$ are the total number of compartments and nodes, respectively. Since $n_\text{a}$ is the total number of arcs, it holds that $n_\text{c} = n_\text{v} + n_\text{a}$ \citep{zocco2023thermodynamical,zocco2022circularity,zocco2024unification}.

Now, since our goal is to design \emph{circular} networks $\mathcal{N}$, we need to define a measure of circularity. We will do so after the following definition.
\begin{definition}[\cite{zocco2024circular}]\label{def:unsMassFlow}
A mass or flow is \emph{finite-time sustainable} if either it exits a nonrenewable reservoir or it enters a landfill, an incinerator, or the natural environment as a pollutant.
\end{definition}
The locations mentioned in Definition \ref{def:unsMassFlow}, i.e., reservoirs, landfills, incinerators, and the environment, are thermodynamic compartments $c^k_{i,j} \in \mathcal{N}$. We can now define the circularity of $\mathcal{N}$.  
\begin{definition}[\cite{zocco2024circular}]\label{def:circularity}
The instantaneous circularity $\lambda(\mathcal{N}; t)$ is defined as
\begin{equation}\label{eq:circularity}
\lambda(\mathcal{N}; t) = - \left(m_{\textup{f},\textup{b}}(t) + \Delta\dot{m}_{\textup{f},\textup{c}}(t)\right).
\end{equation}
\end{definition}
In (\ref{eq:circularity}), $m_{\text{f},\text{b}}(t)$ is the net finite-time sustainable mass transported in batches (e.g., solids transported on trucks), $\dot{m}_{\text{f},\text{c}}(t)$ is the net finite-time sustainable flow transported continuously (e.g., fluids transported through pipes), and $\Delta > 0$ is a constant interval of time introduced as a multiplying factor in order to convert the flow $\dot{m}_{\text{f},\text{c}}(t)$ into a mass, and hence, to make the sum with $m_{\text{f},\text{b}}(t)$ physically consistent. The choice of $\Delta$ is arbitrary, but its value must be kept the same for any calculation of $\lambda(\mathcal{N}; t)$ in order to make comparisons. 

\revision{\textbf{Remark:} The circularity $\lambda(\mathcal{N}; t)$ takes into account the mass transfers taking place in a system, but only those that are finite-time sustainable as defined in Definition \ref{def:unsMassFlow}. We derived the notion of circularity (\ref{eq:circularity}) by formulating the Ellen MacArthur Foundation's definition of circular economy \citep{EllenMacArthurFound} in terms of fundamental physics principles, i.e., in terms of mass transfers. The explicit dependence on time indicates that circularity is a time-varying quantity since it is directly affected by the dynamics of the mass transfers in the system. The negative sign before the parentheses in the right-hand side of the equality in (\ref{eq:circularity}) was added to state the design for systems circularity as a maximization problem. Indeed, with the minus, reducing the extraction of finite resources, e.g., gold or copper, yields a reduction of the quantity inside the parentheses, and hence, an increase of $\lambda(\mathcal{N}; t)$. A good measure of circularity must increase with reductions of raw material mining coherently with the Ellen MacArthur Foundation's definition \citep{EllenMacArthurFound}.     
}

\section{Features of CiRL}\label{sec:Features}
This library has the following four features that make it significantly different from existing libraries of RL environments.  
\begin{enumerate}
\item{\textbf{Focused on material circularity:} To the best of our knowledge, CiRL is the first RL library whose goal is to optimize $\lambda(\mathcal{N}; t)$ (\ref{def:circularity}). The intersection between RL and CE is possible thanks to the formalism of TMNs (Definition \ref{def:TMN}) \citep{zocco2023thermodynamical,zocco2024unification,zocco2022circularity,zocco2024synchronized,zocco2024circular}, where one or more thermodynamic compartments are controlled by RL agents that aim to maximize $\lambda(\mathcal{N}; t)$.}
\item{\textbf{Developed in Google Colaboratory:} CiRL is mainly a collection of Python notebooks written and executed on Google Colaboratory (aka Colab) \citep{ColabWebpage}, and hence, it is more user-friendly than libraries developed with plain programming scripts, e.g., Stable-Baselines3 (SB3) \citep{raffin2021stable}, Gymnasium \citep{towers2024gymnasium}, and TorchRL \citep{boutorchrl}. The extensive use of Python notebooks is in common with the FinRL library \citep{liu2020finrl}. This makes reinforcement learning more accessible to circular economy researchers and engineers whose programming is not the main skill. Since Google maintains the hardware and drivers of Colab, the user needs only to focus on high-level Python programming. To save the training results and models, Google Drive is the default storage location in CiRL. Colab has a free plan suitable for small-scale experiments \revision{and the NVIDIA Tesla T4 is the GPU available for free}; subscription and pay-as-you-go plans are both available if more computing resources are needed, e.g., a more powerful GPU.}
\item{\textbf{Based on SB3:} SB3 is a state-of-the-art library of reinforcement learning algorithms actively maintained and updated \citep{raffin2021stable}. CiRL takes the RL algorithms directly from SB3 and proposes a first collection of environments for the circulation of solids and fluid materials. Specifically, the thermodynamic compartments of (\ref{def:TMNset}) for the circularity of \emph{solid} materials currently implemented in CiRL are a robot for waste sorting, a truck for transportation, and an incinerator for waste removal and energy generation \citep{zocco2024unification}; the only \emph{fluid} currently addressed in CiRL is the atmospheric carbon dioxide because it is one of the main causes of climate instability \citep{NASA-ClimateChange}. Microalgae systems are the thermodynamic compartments considered in CiRL for carbon removal \citep{bernard2016modelling,zocco2025circular}.}
\item{\textbf{Environments in state-space form:} CiRL leverages the robot environments available in Gymnasium Robotics  \citep{towers2024gymnasium}, but it also implements new environments specific for circular economy designs, e.g., an incinerator, an algal cultivation. The new environments are written in the state-space representation, which is the standard form used in systems and control theory for the analysis of systems and design of controllers \citep{goodwin2001control}. Hence, systems and control researchers and engineers can easily interpret CiRL to design RL controllers.}
\end{enumerate}

\section{Compartments \revision{and Compartmental Networks}}\label{sec:Compartments}
In this section, we describe the open-source environments currently created in CiRL. The environments and the tested SB3 algorithms are summarized in Table \ref{tab:summaryEnvsAlgs} along with their roles as network compartments. The RL algorithms are: the A2C \citep{mnih2016asynchronous}, the augmented random search (ARS) \citep{NEURIPS2018_7634ea65}, the deep deterministic policy gradient (DDPG) \citep{lillicrap2019continuouscontroldeepreinforcement}, the proximal policy optimization (PPO) \citep{schulman2017proximal},  the soft actor-critic (SAC) \citep{haarnoja2018soft}, the truncated quantile critics (TQC) \citep{kuznetsov2020controlling}, and the twin delayed DDPG (TD3) \citep{fujimoto2018addressing}. The combination of one of these algorithms with the hindsight experience replay (HER) \citep{andrychowicz2017hindsight} is indicated as ``algorithm-HER'', e.g., ``DDPG-HER'' in the case of DDPG with HER. 
\begin{table*}
\small
\centering
\caption{Summary of current CiRL environments, corresponding compartment $c^k_{i,j}$, tested SB3 algorithms, and corresponding section of the paper covering the details.}
\label{tab:summaryEnvsAlgs}
\begin{tabular}{ccccc} 
Env. name & Comp. & Implementation & SB3 algorithms & Section\\ 
\hline
\emph{Reacher} & \makecell{$c^2_{2,2}$ in \\ Fig. \ref{fig:compDiagraphs_solids}} & \makecell{Same as in \\ Gymnasium} & \makecell{A2C, ARS, DDPG, \\ PPO, SAC} & \ref{subsub:reacher}\\
\hline
\emph{FetchWasteSorting} & \makecell{$c^2_{2,2}$ in \\ Fig. \ref{fig:compDiagraphs_solids}} & \makecell{Modification of \\ \emph{Fetch-PickAndPlace} of \\ Gymnasium Robotics} & \makecell{DDPG-HER, SAC-HER, \\ TD3-HER, TQC-HER} & \ref{subsub:wasteSort}\\
\hline
\emph{TransportTruck} & \makecell{$c^6_{2,3}$ in \\ Fig. \ref{fig:compDiagraphs_solids}} & \makecell{Our implementation \\ in state-space form} & \makecell{A2C, ARS, DDPG, \\ PPO, SAC} & \ref{subsub:truck}\\
\hline
\emph{Incinerator} & \makecell{$c^3_{3,3}$ in \\ Fig. \ref{fig:compDiagraphs_solids}} & \makecell{Our implementation \\ in state-space form} & \makecell{A2C, ARS, DDPG, \\ PPO, SAC} & \ref{subsub:incinerator}\\
\hline
\emph{CO2MicroalgaeMonod} & \makecell{$c^3_{3,3}$ in \\ Fig. \ref{fig:CompDigraph_CO2}} & \makecell{Our implementation \\ in state-space form} & \makecell{A2C, ARS, DDPG, \\ PPO, SAC} & \ref{subsub:Monod}\\
\hline
\emph{CO2MicroalgaeDroop} & \makecell{$c^3_{3,3}$ in \\ Fig. \ref{fig:CompDigraph_CO2}} & \makecell{Our implementation \\ in state-space form} & \makecell{A2C, ARS, DDPG, \\ PPO, SAC} & \ref{subsub:Droop}\\
\hline
\emph{\revision{CarboNet}} & \revision{15 comp.} & \revision{\makecell{Our implementation \\ in state-space form}} & \revision{\makecell{A2C, ARS, DDPG, \\ PPO, SAC}} & \revision{\ref{subsub:CarboNet}}\\
\hline
\emph{\revision{CopperNet}} & \revision{17 comp.} & \revision{\makecell{Our implementation \\ in state-space form}} & \revision{\makecell{A2C, ARS, DDPG, \\ PPO, SAC}} & \revision{\ref{subsub:CopperNet}}\\
\hline
\end{tabular}
\end{table*}

\subsection{Compartments for Circularity of Solids}
\begin{figure}
\includegraphics[width=0.47\textwidth]{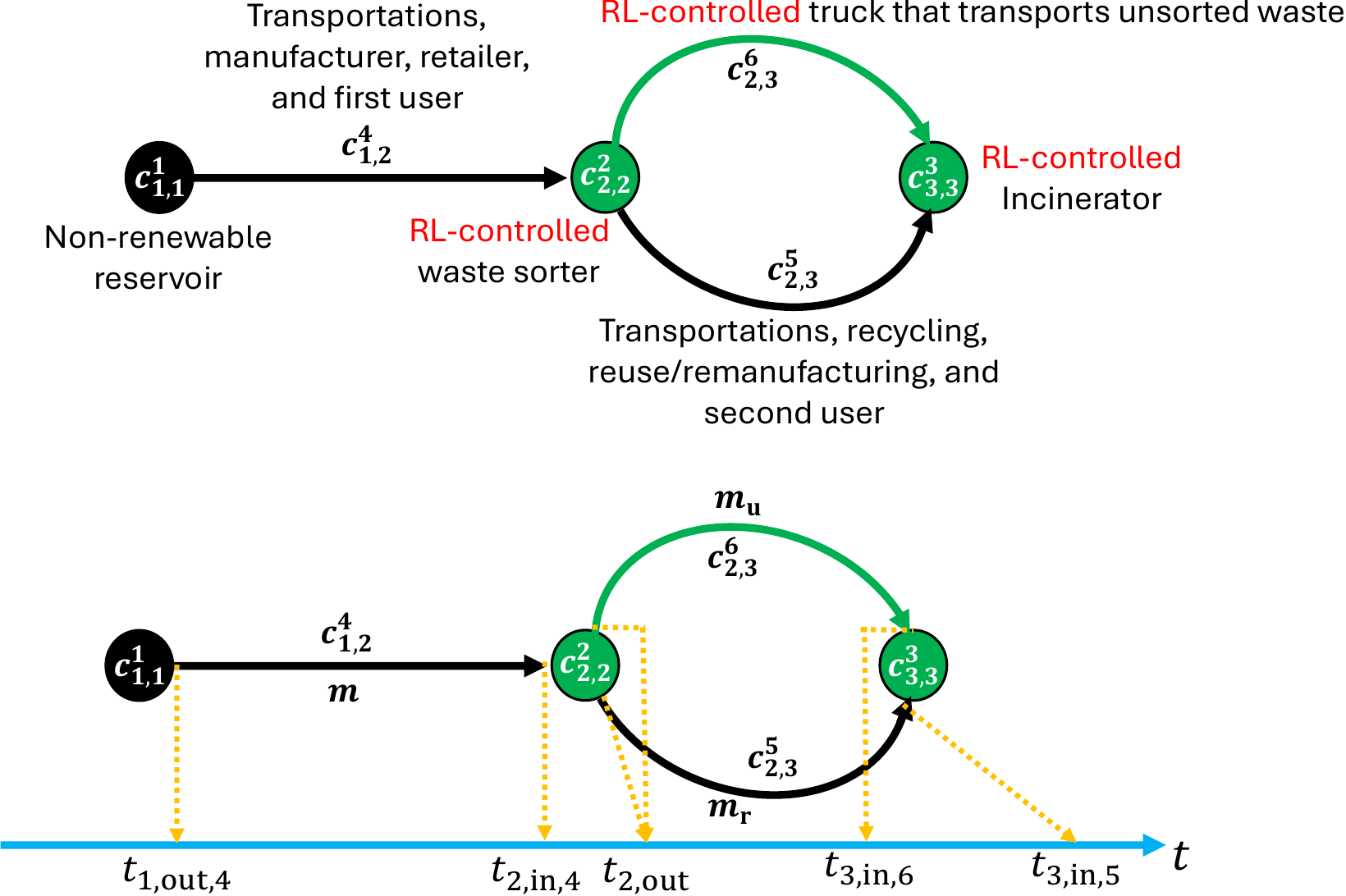}
\centering
\caption{Compartmental diagraph of $\mathcal{N}_\text{s}$ (\ref{eq:Nsolids}). At the top, the description of each compartment; at the bottom, the masses and times are indicated with respect to the temporal axis. In green the considered RL environments.}
\label{fig:compDiagraphs_solids}
\end{figure}

This section considers the material network
\begin{equation}\label{eq:Nsolids}
\mathcal{N}_\text{s} = \{c^1_{1,1}, c^2_{2,2}, c^3_{3,3}, c^4_{1,2}, c^5_{2,3}, c^6_{2,3}\}
\end{equation}
to process solid materials. Its compartmental diagraph is shown in Fig. \ref{fig:compDiagraphs_solids}. Specifically, the diagraph at the top explains the life-cycle stages of the material from extraction (in $c^1_{1,1}$), to robotic waste sorting after first use (in $c^2_{2,2}$), and finally to incineration (in $c^3_{3,3}$). The material exits the waste sorting facility ($c^2_{2,2}$) following two paths depending whether it was successfully sorted or not. The unsorted material is sent directly to incineration using a truck $(c^6_{2,3})$, whereas the sorted material is recycled and used a second time before being sent to incineration $(c^5_{2,3})$. As a result, the mass of unsorted material, which is indicated with $m_{\text{u}}$ in the digraph at the bottom, reaches the incinerator well before the mass of recycled material, namely, $m_{\text{r}}$. Thus, as indicated in the digraph at the bottom of Fig. \ref{fig:compDiagraphs_solids}, $t_{3,\text{in},6} << t_{3,\text{in},5}$, where $t_{i,\text{in},j}$ indicates the time at which the material enters the compartment $i$ from the compartment $j$. Similarly, the notation $t_{i,\text{out},j}$ indicates the time at which the material exits the compartment $i$ for the compartment $j$. As shown at the bottom of Fig. \ref{fig:compDiagraphs_solids}, we assume that the sorted and unsorted batches of waste exit the sorting facility at the same time, and hence, $t_{2,\text{out},5} = t_{2,\text{out},6} = t_{2,\text{out}}$. The yellow arrows in the digraph at the bottom of Fig. \ref{fig:compDiagraphs_solids} specify the correspondence between space and time, that is, at which time the material is at a certain point, e.g., the material exits $c^1_{1,1}$ at $t_{1,\text{out},4}$. 
The nodes and the arc in green are the RL environments considered in this section. Specifically, Sections \ref{subsub:reacher} and \ref{subsub:wasteSort} cover robotic waste sorting, Section \ref{subsub:truck} covers the truck transporting the unsorted waste, and Section \ref{subsub:incinerator} covers the incinerator.

In addition, the digraph at the bottom of Fig. \ref{fig:compDiagraphs_solids} indicates the masses moving between the nodes as weights associated to the arcs. Specifically, $m$ is extracted from the reservoir and, after first use in $c^4_{1,2}$, it enters the waste sorter (when $t$ = $t_{2,\text{in},4}$). For the mass conservation principle, it holds that 
\begin{equation}
m = m_{\text{r}} + m_{\text{u}}. 
\end{equation}

\begin{figure}
\includegraphics[width=0.47\textwidth]{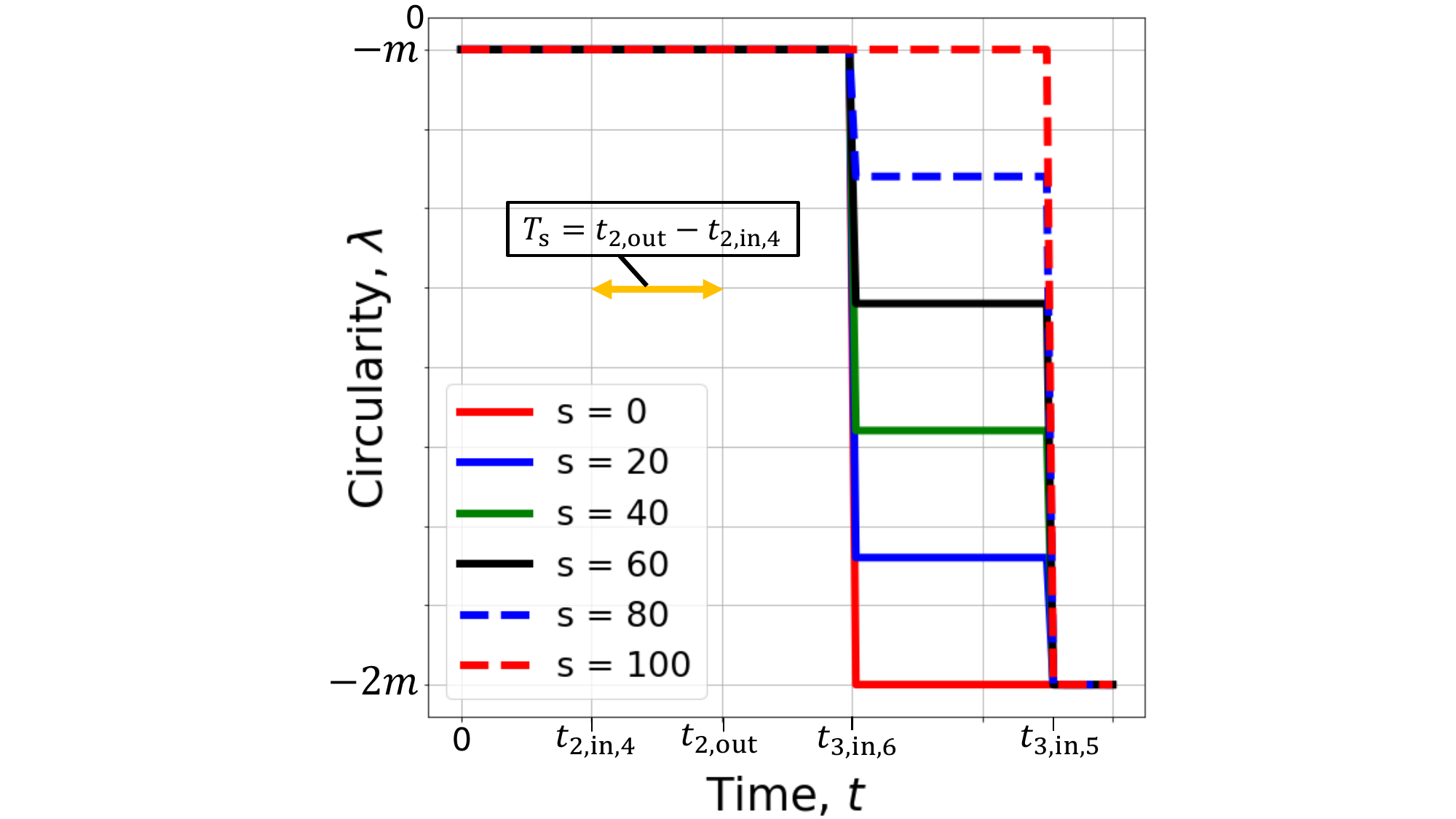}
\centering
\caption{Circularity (\ref{eq:lambdaOftsTs}) of $\mathcal{N}_\text{s}$ as a function of $t$, $s$, and $T_\text{s}$.}
\label{fig:lambda_plot}
\end{figure}

Now, let us see how the performance of an RL-controlled compartment affects the circularity $\lambda(\mathcal{N}_\text{s})$ (Definition \ref{def:circularity}), with $\dot{m}_{\text{f},\text{c}}$ = 0 because no flows of fluids occur in $\mathcal{N}_\text{s}$. In particular, let us consider the robotic waste sorter ($c^2_{2,2}$). By indicating the percentage success of sorting with $s \in [0,100]$, we have that  
\begin{equation}
m_{\text{u}} = m\left(1-\frac{s}{100}\right)
\end{equation}
and
\begin{equation}
m_{\text{r}} = m \frac{s}{100}.
\end{equation}
Thus, the dynamics of circularity is
\begin{equation}\label{eq:lambdaOftsTs}
\lambda(\mathcal{N}_{\text{s}}; t, s, T_\text{s}) =
\begin{cases}
-m, \quad 0 \leq t < t_{3,\text{in},6}(T_\text{s}) \\
-(m + m_{\text{u}}), \quad t_{3,\text{in},6}(T_\text{s}) \leq t < t_{3,\text{in},5}(T_\text{s}) \\
-2m, \quad t \geq t_{3,\text{in},5}(T_\text{s}), 
\end{cases}
\end{equation}
where $T_\text{s}$ is the time required for sorting the material. Figure \ref{fig:lambda_plot} is a graphical representation of (\ref{eq:lambdaOftsTs}). As shown, an increase (decrease) of $T_\text{s}$ delays (anticipates) the subsequent events that occur at $t_{2,\text{out}}$, $t_{3,\text{in},6}$, and $t_{3,\text{in},5}$.      
The success of sorting $s$ affects the magnitude of $\lambda$. For example, $s$ = 0 yields that all the waste material enters the incinerator at $t_{3,\text{in},6}$, i.e., $\lambda_\text{a}$ = $\lambda(\mathcal{N}_{\text{s}}; t_{3,\text{in},6}, 0, T_\text{s})$ = $-2m$. In contrast, $s$ = 100 increases circularity by extending the life-cycle of the material, i.e., $\lambda(\mathcal{N}_{\text{s}}; t_{3,\text{in},6}, 100, T_\text{s})$ = $-m > \lambda_\text{a}$. While all the material is inside the incinerator for $t \geq t_{3,\text{in},5}$ regardless of $s$, i.e.,  
\begin{equation}
\lambda(\mathcal{N}_{\text{s}}; t_{3,\text{in},5}, s, T_\text{s}) = -2m, \quad \forall s,
\end{equation}
the recycled mass $m_\text{r}$ has extended the life-cycle of the amount $\nu$ = $t_{3,\text{in},5}$ - $t_{3,\text{in},6}$ if $s > 0$, which yields an increase of $m_\text{r}$ in $\lambda$ for $t_{3,\text{in},6} \leq t < t_{3,\text{in},5}$ ($\lambda$ is independent of $s$ for $t < t_{3,\text{in},6}$ and for $t \geq t_{3,\text{in},5}$). Extending the life-cycle of materials is a key principle of circular economy \citep{potting2017circular}. 

Now, note that both $T_\text{s}$ and $s$ are affected by the performance of the RL-controlled robotic sorter. For example, a fast and accurate RL-controller yields a small $T_\text{s}$ and a high $s$. In contrast, a fast and inaccurate sorting yields a small $T_\text{s}$ and a low $s$. Hence, \emph{the performance of the RL controller affects the circularity $\lambda(\mathcal{N}_\text{s})$ through $s$ and $T_\text{s}$} (see Fig. \ref{fig:lambda_plot}).

\subsubsection{Reacher}\label{subsub:reacher}
The \emph{Reacher} environment \citep{towers2024gymnasium} consists of a robot with two revolute joints rotating around a direction normal to the plane, it has two degrees of freedom, its actions are the torques applied to the revolute joints, and the goal of the robot is to reach a target point generated at a random location. It is a simplification of the pick-and-place task needed to perform waste sorting. Its simplicity enable us to illustrate how the percentage success $s$ is affected by the RL controller performance, which in turns affects $\lambda$ as shown in Fig. \ref{fig:lambda_plot}. We tested the five SB3 algorithms reported in Table \ref{tab:results-Reacher}, where
\begin{equation}
\zeta = r_\text{e} - r_\text{s},
\end{equation}
with $r_\text{e}$ and $r_\text{s}$ denoting the mean reward over 100 episodes at the end and at the start of the training, respectively\revision{, and where the sample efficiency is quantified via
\begin{equation}
\psi = \frac{\zeta}{n_\text{ts}},
\end{equation}
with $n_\text{ts}$ the number of training time steps. Thus, $\psi$ is the average increase of the reward per time step, i.e., per interaction between the agent and the environment.} The parameter $s$ is calculated by considering as a success an episode in which the end effector reaches a distance from the target smaller than 0.04 m and the joint torques are less than 0.005 Nm. We have run 10,000 episodes and counted the successful episodes. 
\begin{table*}
\centering
\caption{Training time, $r_\text{s}$, $r_\text{e}$, $\zeta$, \revision{$\psi$}, and $s$ for each algorithm with \emph{Reacher}. Trainings were executed for 100,000 time steps on a Colab NVIDIA Tesla T4. In bold values of $\zeta > 0$. \revision{Sample efficiency $\psi$ indicated only for $\zeta > 0$.}}
\label{tab:results-Reacher}
\begin{tabular}{ccccccc} 
 & Tr. time (min:sec) & $r_\text{s}$ & $r_\text{e}$ & $\zeta$ & \revision{$\psi$} & $s$ \\ 
\hline
A2C & 04:37 & -63.4 & -16.6 & \textbf{46.8} & \revision{$4.7\times 10^{-4}$} & 17.8 \\
ARS & 01:00 & -12.1 & -11.0 & \textbf{1.1} & \revision{$1.1\times 10^{-5}$} & 2.8 \\
DDPG & 13:43 & -60.2 & -5.5 & \textbf{54.7} & \revision{$5.5\times 10^{-4}$} & 72.3 \\
PPO & 03:59 & -60.2 & -37.3 & \textbf{22.9} & \revision{$2.3
\times 10^{-4}$} & 20.6 \\
SAC & 22:23 & -45.1 & -8.3 & \textbf{36.8} & \revision{$3.7\times 10^{-4}$} & 51.7\\
\hline
\end{tabular}
\end{table*}

\subsubsection{Fetch Waste Sorting}\label{subsub:wasteSort}
The \emph{Reacher} environment is a very simplified model of robotic waste sorting. A more realistic simulator is the one based on the Gymnasium Robotics environment \emph{Fetch-PickAndPlace}, which is a 7-DOF manipulator that performs pick-and-place with a two-finger gripper \citep{plappert2018multi,towers2024gymnasium}. The RL agent controls the opening and closure of the fingers along with the motion of the robot joints. To resemble a waste sorting tasks we modified the Gymnasium environment by constraining the target to be on the table instead of in the air and by locating the target at a fixed point instead of being randomly generated since the coordinates of the point where to place a waste item are usually known and fixed, e.g., the position of the bin. Hence, this environment is referred to as \emph{FetchWasteSorting}. With this environment, we tested the SB3 algorithms whose performance are reported in Table \ref{tab:results-Fetch} \revision{and in Figs. \ref{fig:FetchDDPG}-\ref{fig:FetchTQC}.}
\begin{table*}
\centering
\caption{Training time, $r_\text{s}$, $r_\text{e}$, \revision{$\psi$}, and $\zeta$ for each algorithm with \emph{FetchWasteSorting}. Trainings were executed for 150,000 time steps on a Colab NVIDIA Tesla T4. In bold values of $\zeta > 0$. \revision{Sample efficiency $\psi$ indicated only for $\zeta > 0$.}}
\label{tab:results-Fetch}
\begin{tabular}{cccccc} 
 & Tr. time (min:sec) & $r_\text{s}$ & $r_\text{e}$ & $\zeta$ & \revision{$\psi$}\\ 
\hline
DDPG-HER & 50:06 & -50.25 & -50 & \textbf{0.25} & \revision{$1.6 \times 10^{-6}$} \\
SAC-HER & 63:43 & -50.25 & -12.67 & \textbf{37.58} & \revision{$2.5 \times 10^{-4}$} \\
TD3-HER & 45:12 & -50.25 & -50 & \textbf{0.25} & \revision{$1.6 \times 10^{-6}$}\\
TQC-HER & 65:19 & -50.25 & -11.45 & \textbf{38.80} & \revision{$2.6 \times 10^{-4}$}\\
\hline
\end{tabular}
\end{table*}

\subsubsection{Transport Truck}\label{subsub:truck}
The transport truck (compartment $c^6_{2,3}$) moves the unsorted waste of mass $m_\text{u}$ from the sorting facility ($c^2_{2,2}$) to the incinerator ($c^3_{3,3}$) as shown in Fig. \ref{fig:compDiagraphs_solids}. Currently, the model of the truck in CiRL consists of a particle of mass $m_\text{tot} = m_\text{truck} + m_\text{u}$, where $m_\text{truck}$ is the mass of the empty truck. As demonstrated by \cite{zocco2023thermodynamical}, the first law of thermodynamics leads to the Lagrange's equations, which in turns yield the second Newton's law of motion \citep{zocco2024circular}. Considering that the particle-truck is subject to the traction force $F$, the state-space form of the system is
\begin{equation}
\dot{\bm{x}} = 
\begin{bmatrix}
\dot{x}_1 \\
\dot{x}_2 \\
\end{bmatrix}
=
\begin{bmatrix}
x_2 \\
\frac{F}{m_\text{tot}} \\
\end{bmatrix},
\end{equation}
where $x_1$ and $x_2$ are the position and the speed of the truck, respectively. The observations of this environment correspond to the state variables $x_1$ and $x_2$, while $F$ is the action. The reward function is written as 
\begin{equation}
r = - \left[\left(x_\text{inc} - x_1\right)^2 + 0.1 x_2^2 + 0.001 F^2\right],
\end{equation}
and hence, the RL controller trained successfully generates a force $F$ that brings the truck to the incinerator located at $x_\text{inc}$ and stops it there. This condition corresponds to the maximum reward $r_\text{max} = 0$. The performance of a single run of the tested RL algorithms is reported in Table \ref{tab:results-Truck}. \revision{It is worth noticing that the proposed model is simplified and does not consider elements that could be important in practice such as the friction, the aerodynamic drag, or real-world terrain variability. For transportation systems based on railway vehicles, for example, the model presented by \cite{malvezzi2013tool} includes different motion resistance sources.}
\begin{table*}
\centering
\caption{Training time, $r_\text{s}$, $r_\text{e}$, $\zeta$, and \revision{$\psi$} for each algorithm with \emph{TransportTruck}. Trainings were executed for 400,000 time steps on a Colab NVIDIA Tesla T4. In bold values of $\zeta > 0$. \revision{Sample efficiency $\psi$ indicated only for $\zeta > 0$.}}
\label{tab:results-Truck}
\begin{tabular}{cccccc} 
 & Tr. time (min:sec) & $r_\text{s}$ & $r_\text{e}$ & $\zeta$ & \revision{$\psi$}\\ 
\hline
A2C & 18:10 & $-1.0 \times 10^{11}$ & $-1.2 \times 10^{8}$ & $\bm{1.0 \times 10^{11}}$ & \revision{$2.5 \times 10^{5}$}\\
ARS & 02:56 & $-4.0 \times 10^{8}$ & $-7.1 \times 10^{8}$ & $-3.1 \times 10^{8}$ & -- \\
DDPG & 50:42 & $-1.7 \times 10^{11}$ & $-2.0 \times 10^{11}$ & $-3.3 \times 10^{10}$ & -- \\
PPO & 14:34 & $-1.03 \times 10^{11}$ & $-1.02 \times 10^{11}$ & $\bm{8.6 \times 10^{8}}$ & \revision{$2.1 \times 10^{3}$}\\
SAC & 1:28:18 & $-9.8 \times 10^{10}$ & $-1.3 \times 10^{11}$ & $-3.4 \times 10^{10}$ & --\\
\hline
\end{tabular}
\end{table*}

\subsubsection{Incinerator}\label{subsub:incinerator}
The compartment $c^3_{3,3}$ in Fig. \ref{fig:compDiagraphs_solids} is the incinerator, which processes end-of-life products and materials. From a circular economy perspective, it is among the least circular options because it destroys the functionality of products and their components \citep{potting2017circular}. A complete dynamical model of an incinerator was developed by \cite{magnanelli2020dynamic} as a result of mass and energy balances. In this first version of CiRL, we consider only the wastebed and the freeboard sub-compartments of \cite{magnanelli2020dynamic}. This yields the following system of six ordinary differential equations in the state-space form and with state vector $\bm{x} = [M, M_\text{char}, T_\text{w}, M_\text{g,w}, M_\text{g,f}, T_\text{g}]^\top = [x_1,x_2,x_3,x_4,x_5,x_6]^\top$, where $M$ is the mass of waste in the wastebed, $M_\text{char}$ is the mass of char in the wastebed, $T_\text{w}$ is the temperature in the wastebed, $M_\text{g,w}$ is the mass of the gasses in the wastebed, $M_\text{g,f}$ is the mass of the gasses in the freeboard, and $T_\text{g}$ is the temperature of the gasses in the freeboard.
\begin{equation}
\dot{\bm{x}} = 
\begin{bmatrix}
\dot{x}_1 \\
\dot{x}_2 \\
\dot{x}_3 \\
\dot{x}_4 \\
\dot{x}_5 \\
\dot{x}_6 
\end{bmatrix}
=
\begin{bmatrix}
F_\text{in} - F_\text{out} - R_\text{w} \\
-F_\text{char,out} + P_\text{char} - R_\text{char} \\
\frac{F_\text{in}c_\text{p,w}x_3 + F_\text{aI}c_\text{p,g}\left(T_\text{aI} - x_3\right) + Q}{c_\text{p,w}x_1 + c_\text{p,char}x_2 + c_\text{p,m}M_\text{grate}} \\
F_\text{aI} - F_\text{g,w,out} + R_\text{g} \\
F_\text{g,w,out} - F_\text{g,out} + F_\text{aII} \\
\frac{F_\text{g,w,out} c_\text{p,g}\left(x_3 - x_6\right) + F_\text{aII} c_\text{p,g} \left(T_\text{aII} - x_6\right) + Q_\text{g} - Q_\text{ext}}{c_\text{p,g}x_5+c_\text{p,m}M_\text{fb}}
\end{bmatrix},
\end{equation}
where $F_\text{in}$ is the input mass flow rate of waste to the incinerator, $F_\text{out}$ is the output flow of ashes from the incinerator, $R_\text{w}$ is the consumption rate of waste via pyrolysis and combustion, $F_\text{char,out}$ is the output mass flow rate of char, $P_\text{char}$ is the production rate of char due to pyrolysis, $R_\text{char}$ is the consumption rate due to combustion, $c_\text{p,w}$ and $c_\text{p,g}$ are the specific heat capacities of waste and air, $F_\text{aI}$ and $T_\text{aI}$ are the input mass flow rate and the temperature of the primary air entering the wastebed, $Q$ is the heat released by the waste conversion (exothermic reaction), $c_\text{p,char}$ is the specific heat capacity of the char, $c_\text{p,m}$ is the specific heat capacity of the metal making up the grate and the freeboard, $M_\text{grate}$ is the grate mass, $F_\text{g,w,out}$ is the mass flow rate of gases that exits the wastebed and enters the freeboard, $R_\text{g}$ is the rate of production of gasses in the wastebed, $F_\text{g,out}$ is the mass flow rate of gasses exiting the freeboard (i.e., the incinerator), $F_\text{aII}$ and $T_\text{aII}$ are the mass flow rate and the temperature of the secondary air entering the freeboard, $Q_\text{g}$ is the reaction heat in the freeboard due to combustion, $M_\text{fb}$ is the mass of the freeboard, and $Q_\text{ext}$ is the heat exchanged between the freeboard and the surrounding. The latter is the action in this environment and it is generated by the RL controller to regulate the temperature in the freeboard $x_6$ to the desired value, namely, $T_\text{g,d}$. Hence, the reward function for this environment is
\begin{equation}
r_\text{i} = - (T_\text{g,d} - x_6)^2.  
\end{equation}
The performance of a single run of the tested RL algorithms
on the incinerator environment is reported in Table \ref{tab:results-Incinerator} \revision{and in Figs. \ref{fig:incineratorA2C}-\ref{fig:incineratorPPO}}.
\begin{table*}
\centering
\caption{Training time, $r_\text{s}$, $r_\text{e}$, $\zeta$, and \revision{$\psi$} for each algorithm with \emph{Incinerator}. Trainings were executed for 200,000 time steps on a Colab NVIDIA Tesla T4. In bold values of $\zeta > 0$. \revision{Sample efficiency $\psi$ indicated only for $\zeta > 0$.}}
\label{tab:results-Incinerator}
\begin{tabular}{ccccccc} 
 & Tr. time (min:sec) & $r_\text{s}$ & $r_\text{e}$ & $\zeta$ & \revision{$\psi$}\\ 
\hline
A2C & 08:25 & $-3.40 \times 10^5$ & $-3.79 \times 10^5$ & $-3.9 \times 10^4$ & --\\
ARS & 01:28 & $-3.65 \times 10^5$ & $-3.53 \times 10^5$ & $\bm{1.21 \times 10^4}$ & \revision{$6.0 \times 10^{-2}$}\\
DDPG & 24:50 & $-3.23 \times 10^5$ & $-3.44 \times 10^5$ & $-2.07 \times 10^4$ & --\\
PPO & 07:30 & $-3.51 \times 10^5$ & $-3.58 \times 10^5$ & $-0.71 \times 10^4$ & --\\
SAC & 43:20 & $-3.37 \times 10^5$ & $-3.41 \times 10^5$ & $-0.40 \times 10^4$ & --\\
\hline
\end{tabular}
\end{table*}

\subsection{Compartments for Circularity of $\text{CO}_2$}
\begin{figure}
\includegraphics[width=0.47\textwidth]{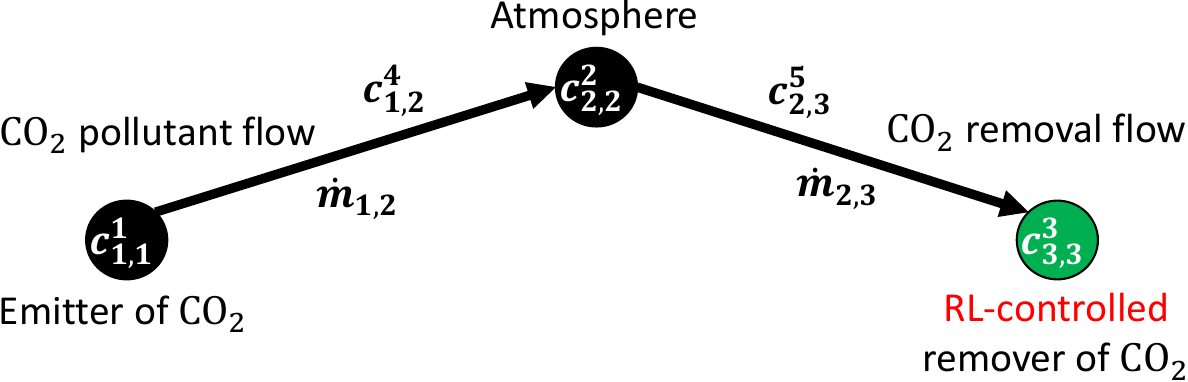}
\centering
\caption{Compartmental diagraph of $\mathcal{N}_{\text{nz}}$ (\ref{eq:netForNZ}). In green the considered RL environments.}
\label{fig:CompDigraph_CO2}
\end{figure}

In this section, we address the \emph{circulation} of atmospheric carbon dioxide since its accumulation is one of the main causes of climate change \citep{NASA-ClimateChange}. Governments usually refer to the need of carbon dioxide circulation as the ``net-zero target'', which is defined by \cite{NZ-DefUniOx}, e.g., the EU \citep{EU-NZ}, the UK \citep{UK-NZ}, the US \citep{US-NZ}, China \citep{China-NZ}, and India \citep{India-NZ}. Using the TMN formalism, the net-zero problem can be represented as
\begin{equation}\label{eq:netForNZ}
\mathcal{N}_{\text{nz}} = \{c^1_{1,1}, c^2_{2,2}, c^3_{3,3}, c^4_{1,2}, c^5_{2,3}\}
\end{equation}
and it is depicted in Fig. \ref{fig:CompDigraph_CO2}. 
The net-zero problem can be formulated as the design of $\mathcal{N}_{\text{nz}}$ involving one or more emitters and removers of $\text{CO}_2$ such that the accumulation in the atmosphere ($c^2_{2,2}$) is zero, i.e., a system whose circularity is
\begin{equation}\label{eq:lambdaNZ}
\begin{gathered}
\lambda(\mathcal{N}_{\text{nz}}; t) = - \Delta\dot{m}_{\text{f,c}}(t) = - \Delta\left(\dot{m}_{1,2}(t) - \dot{m}_{2,3}(t)\right) = 0 \\
\Rightarrow \quad \dot{m}_{1,2}(t) = \dot{m}_{2,3}(t),
\end{gathered}
\end{equation}
where $\dot{m}_{1,2}(t)$ is the continuous flow of $\text{CO}_2$ produced by the emitter compartment and $\dot{m}_{2,3}(t)$ is the continuous flow of $\text{CO}_2$ absorbed by the remover compartment (Fig. \ref{fig:CompDigraph_CO2}).
Equation (\ref{eq:lambdaNZ}) shows that the net-zero condition corresponds to $\lambda(\mathcal{N}_{\text{nz}}; t) = 0$, i.e., when circularity is maximum (Definition \ref{def:circularity}). This is achieved when the emitted $\text{CO}_2$ equals the removed $\text{CO}_2$. The compartments considered in this section are highlighted in green in Fig. \ref{fig:CompDigraph_CO2}.

Currently, CiRL considers microalgae cultivations as the remover compartment \citep{goswami2024advances}. Microalgae are promising for carbon removal because they are a renewable source and, at the end of their life, they can be used as biomass for biofuel production \citep{goswami2024advances,sadvakasova2023microalgae}. They also have shown to absorb more $\text{CO}_2$ than terrestrial plants \citep{zhou2017bio,wang2008co}. The RL controllers currently designed in CiRL aim to improve the ability of microalgae cultivations to absorb $\text{CO}_2$ by regulating the light intensity. Hence, the action space is one-dimensional.

\subsubsection{Monod's Model for Microalgae}\label{subsub:Monod}
One of the simplest models used to describe the dynamics of microalgae cultivations is the Monod's model \citep{bernard2016modelling,zocco2025circular}. It is a state-space model of two states where the growth limiting factors enter the system as multiplying terms in the equation of the growth rate $\mu$. The model implemented in CiRL is detailed in \citep{vatcheva2006experiment,bernard2011hurdles}, while the design of the RL controllers is detailed in \citep{zocco2025circular}. For this paper, we executed a single training with each SB3 algorithm whose performance is reported in Table \ref{tab:results-Monod} \revision{and in Figs. \ref{fig:MonodA2C}-\ref{fig:MonodSAC}}.
\begin{table*}
\centering
\caption{Training time, $r_\text{s}$, $r_\text{e}$, $\zeta$, and \revision{$\psi$} for each algorithm with \emph{CO2MicroalgaeMonod}. Trainings were executed for 200,000 time steps on a Colab NVIDIA Tesla T4. In bold values of $\zeta > 0$. \revision{Sample efficiency $\psi$ indicated only for $\zeta > 0$.}}
\label{tab:results-Monod}
\begin{tabular}{cccccc} 
 & Tr. time (min:sec) & $r_\text{s}$ & $r_\text{e}$ & $\zeta$ & \revision{$\psi$}\\ 
\hline
A2C & 08:10 & 127.7 & 143.1 & \textbf{15.4} & \revision{$7.7 \times 10^{-5}$}\\
ARS & 01:28 & 114.6 & 142.8 & \textbf{28.2} & \revision{$1.4 \times 10^{-4}$}\\
DDPG & 31:34 & 98.3 & 95.4 & -2.9 & --\\
PPO & 07:19 & 116.8 & 135.4 & \textbf{18.6} & \revision{$9.3 \times 10^{-5}$}\\
SAC & 43:22 & 117.4 & 137.4 & \textbf{20.0} & \revision{$1.0 \times 10^{-4}$}\\
\hline
\end{tabular}
\end{table*}

\subsubsection{Droop's Model for Microalgae}\label{subsub:Droop}
The Droop's model can capture the dynamics of microalgae growth more accurately than the Monod's model. With respect to the latter, it has an extra state, which models the internal cell quota \citep{vatcheva2006experiment,bernard2011hurdles}. The Droop's model derives from the application of the mass conservation principle \citep{saccardo2023droop}, and hence, is a thermodynamic compartment of a TMN \citep{zocco2024unification}. The reward function we implemented for the Droop's model is that of the Monod's model, that is \citep{zocco2025circular},
\begin{equation}
\rho_{\text{CO}_2} = K_{\text{CO}_2}\rho,
\end{equation}
where $\rho$ is the uptake rate of nutrients, $K_{\text{CO}_2} \in (0,1)$ is a constant quantifying which fraction of absorbed nutrients is carbon dioxide, and $\rho_{\text{CO}_2}$ is the rate of $\text{CO}_2$ uptake. Thus, the flow $\dot{m}_{2,3}$ in Fig. \ref{fig:CompDigraph_CO2} is given by  
\begin{equation}
\dot{m}_{2,3} = \rho_{\text{CO}_2}.
\end{equation}
The performance of the RL algorithms for a single run with the Droop's model is reported in Table \ref{tab:results-Droop}.
\begin{table*}
\centering
\caption{Training time, $r_\text{s}$, $r_\text{e}$, $\zeta$ and \revision{$\psi$} for each algorithm with \emph{CO2MicroalgaeDroop}. Trainings were executed for 200,000 time steps on a Colab NVIDIA Tesla T4. In bold values of $\zeta > 0$. \revision{Sample efficiency $\psi$ indicated only for $\zeta > 0$.}}
\label{tab:results-Droop}
\begin{tabular}{cccccc} 
 & Tr. time (min:sec) & $r_\text{s}$ & $r_\text{e}$ & $\zeta$ & \revision{$\psi$}\\ 
\hline
A2C & 08:26 & 409.2 & 473.0 & \textbf{63.8} & \revision{$3.2 \times 10^{-4}$}\\
ARS & 01:35 & 461.8 & 461.2 & -0.6 & --\\
DDPG & 26:07 & 437.1 & 471.2 & \textbf{34.1} & \revision{$1.7 \times 10^{-4}$}\\
PPO & 07:07 & 526.5 & 446.3 & -80.2 & --\\
SAC & 44:17 & 343.3 & 459.6 & \textbf{116.3} & \revision{$5.8 \times 10^{-4}$}\\
\hline
\end{tabular}
\end{table*}

\subsection{\revision{Compartmental Networks}}
\revision{This section covers two networks of compartments, namely, CarboNet and CopperNet. The former focuses on the flow of tropospheric carbon dioxide, while the latter on the supply-recovery chain of copper.}

\subsubsection{\revision{CarboNet}}\label{subsub:CarboNet}
\revision{CarboNet is a network with 15 compartments and it addresses the circularity of carbon dioxide. It was recently proposed in \cite{zocco2025carbonet} and controlled via linear quadratic regulators. Here, it is implemented as an SB3 environment to approach its control via reinforcement learning. We set the reward function for this environment as
\begin{equation}
r_\text{ca} = - (x_{1,\text{d}} - x_1)^2,  
\end{equation}
where $x_{1,\text{d}}$ is the desired mass of $\text{CO}_2$ in the troposphere of a target area corresponding to the concentration of $\text{CO}_2$ in the pre-industrial era.
The performance of a single run of the tested RL algorithms
on CarboNet is reported in Table \ref{tab:results-CarboNet} \revision{and in Figs. \ref{fig:CarboNetA2C}-\ref{fig:CarboNetPPO}}.
\begin{table*}
\centering
\caption{\revision{Training time, $r_\text{s}$, $r_\text{e}$, $\zeta$ and \revision{$\psi$} for each algorithm with \emph{CarboNet}. Trainings were executed for 200,000 time steps on a Colab NVIDIA Tesla T4. In bold values of $\zeta > 0$. Sample efficiency $\psi$ indicated only for $\zeta > 0$.}}
\label{tab:results-CarboNet}
\begin{tabular}{cccccc} 
 & Tr. time (min:sec) & $r_\text{s}$ & $r_\text{e}$ & $\zeta$ & \revision{$\psi$}\\ 
\hline
A2C & 07:36 & $-7.70 \times 10^6$ & $-7.06 \times 10^7$ & $-6.29 \times 10^7$ & --\\
ARS & 01:24 & $-3.47 \times 10^7$ & $-2.04 \times 10^6$ & $\bm{3.27 \times 10^7}$ & \revision{163}\\
DDPG & 24:07 & $-4.10 \times 10^7$ & $-7.08 \times 10 ^7$& $-2.97 \times 10^7$ & --\\
PPO & 06:43 & $-1.13 \times 10^7$ & $-1.17 \times 10^7$ & $-3.70 \times 10^5$ & --\\
SAC & 39:27 & $-2.66 \times 10^7$ & $-7.08 \times 10^7$ & $-4.41 \times 10^7$ & --\\
\hline
\end{tabular}
\end{table*}
}

\subsubsection{\revision{CopperNet}}\label{subsub:CopperNet}
\revision{This environment, namely, CopperNet, is made of six linear ordinary differential equations resulting from the mass balance principle applied to six nodes of a supply-recovery chain that extracts, transports, transforms, and use copper. The network architecture is inspired by the material flow analysis in \cite[Fig. 1]{loibl2021current}. The compartmental diagraph of CopperNet is shown in Fig. \ref{fig:CompDigraph_CopperNet},
\begin{figure}
\includegraphics[width=0.47\textwidth]{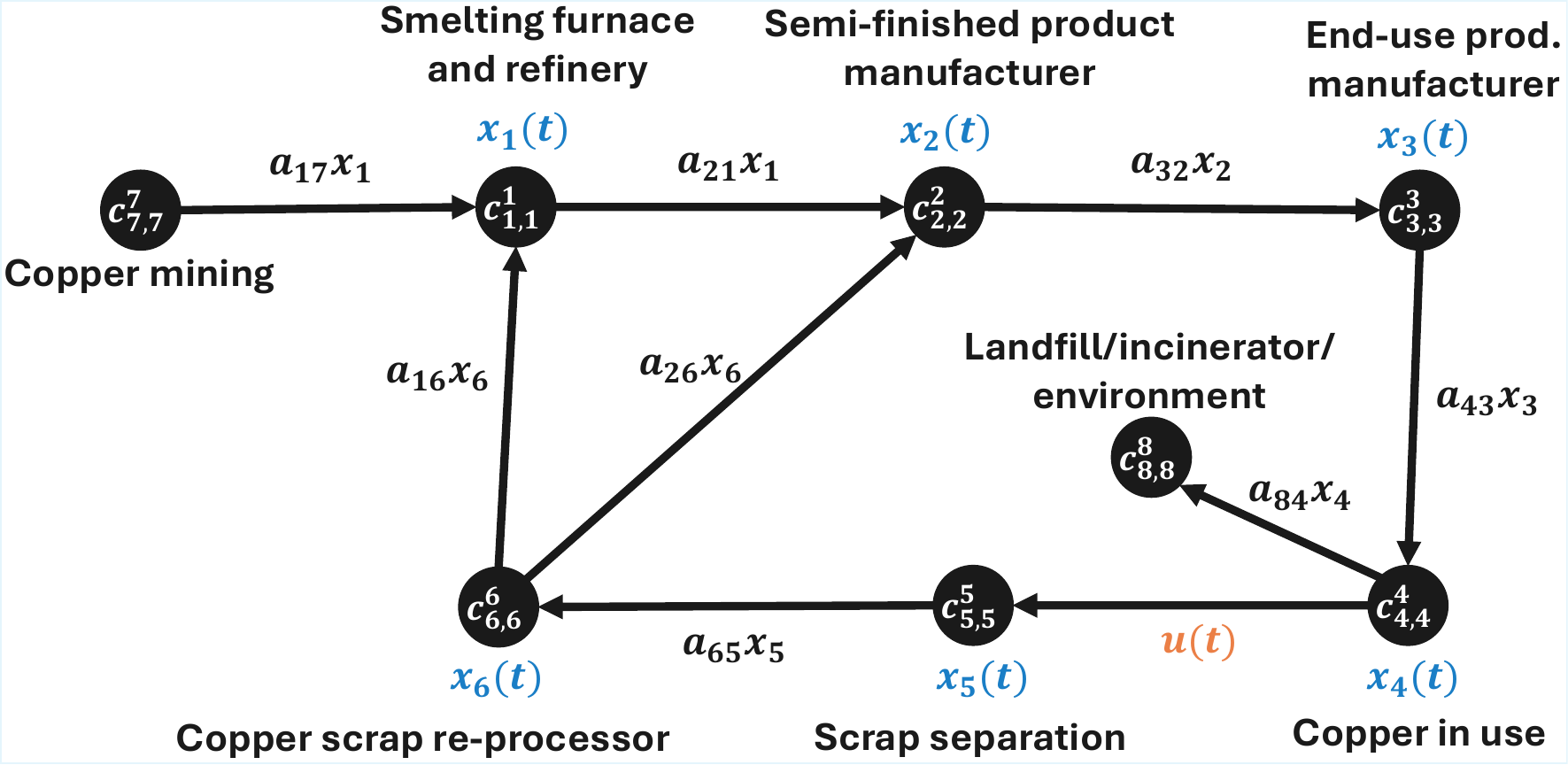}
\centering
\caption{\revision{Compartmental digraph of a supply-recovery network of copper, namely, CopperNet.}}
\label{fig:CompDigraph_CopperNet}
\end{figure}
while its state-space model is given by
\begin{equation}
\dot{\bm{x}} = 
\begin{bmatrix}
\dot{x}_1 \\
\dot{x}_2 \\
\dot{x}_3 \\
\dot{x}_4 \\
\dot{x}_5 \\
\dot{x}_6 
\end{bmatrix}
=
\begin{bmatrix}
a_{17} x_1 + a_{16} x_6 - a_{21} x_1 \\
a_{21} x_1 + a_{26} x_6 - a_{32} x_2 \\
a_{32} x_2 - a_{43} x_3 \\
a_{43} x_3 - a_{84} x_4 - u \\
-a_{65} x_5 + u \\
a_{65} x_5 - a_{16} x_6 - a_{26} x_6
\end{bmatrix},
\end{equation}
where $a_{i,j} \geq$ 0 [1/day] is the rate constant that is proportional to the mass flow rate between the node $i$ and the node $j$ of the digraph (see Fig. \ref{fig:CompDigraph_CopperNet}). The six states $x_i$ [kt], $i = 1, \dots, 6$, are masses located at different stages of the supply-recovery chain of copper as indicated in Fig. \ref{fig:CompDigraph_CopperNet}, while $u$ [kt/day] is the action of the RL agent. Coherently with the notation in Definition \ref{def:TMN}, CopperNet can be compactly represented by the set
\begin{equation}
\begin{gathered}
\mathcal{N}_{\text{c}} = \{c^1_{1,1}, c^2_{2,2}, c^3_{3,3}, c^4_{4,4}, c^5_{5,5}, c^6_{6,6}, c^7_{7,7},c^8_{8,8}, c^{9}_{1,2}, \\ c^{10}_{2,3}, c^{11}_{3,4}, c^{12}_{4,5}, c^{13}_{4,8}, c^{14}_{5,6}, c^{15}_{6,1}, c^{16}_{6,2}, c^{17}_{7,1} \}.
\end{gathered}
\end{equation}
Thus, for $\mathcal{N}_{\text{c}}$ it holds that $n_\text{c}$ = 17, $n_\text{v}$ = 8, and $n_\text{a}$ = 9. By applying Definition \ref{def:circularity}, the circularity of CopperNet is
\begin{equation}
\lambda = -(a_{17}x_1 + a_{84}x_4),
\end{equation}
where the first term on the right-hand side of the equation takes into account the finite-time sustainable extraction of copper, whereas the second term on the right-hand side of the equation takes into account the leaks of copper that end-up in  landfills, incinerators, and the natural environment. 

Finally, we defined the reward function as
\begin{equation}
r_\text{co} = - \sum_{i = 1}^{6} (x_{i,\text{d}} - x_i)^2, 
\end{equation}
where $x_{i,\text{d}}$ is the desired value of the state $x_i$. The performance of a single run of the tested RL algorithms
on CopperNet is reported in Table \ref{tab:results-CopperNet} \revision{and in Figs. \ref{fig:CopperNetA2C}-\ref{fig:CopperNetPPO}}.
\begin{table*}
\centering
\caption{\revision{Training time, $r_\text{s}$, $r_\text{e}$, $\zeta$, and \revision{$\psi$} for each algorithm with \emph{CopperNet}. Trainings were executed for 200,000 time steps on a Colab NVIDIA Tesla T4. In bold values of $\zeta > 0$. Sample efficiency $\psi$ indicated only for $\zeta > 0$.}}
\label{tab:results-CopperNet}
\begin{tabular}{cccccc} 
 & Tr. time (min:sec) & $r_\text{s}$ & $r_\text{e}$ & $\zeta$ & \revision{$\psi$}\\ 
\hline
A2C & 07:26 & $-1.34 \times 10^8$ & $-5.30 \times 10^7$ & $\bm{8.15 \times 10^7}$ & \revision{407}\\
ARS & 01:28 & $-2.81 \times 10^8$ & $-4.71 \times 10^7$ & $\bm{2.34 \times 10^8}$ & \revision{1170}\\
DDPG & 22:56 & $-2.57 \times 10^8$ & $-4.43 \times 10^8$ & $-1.86 \times 10^8$ & --\\
PPO & 06:38 & $-9.32 \times 10^7$ & $-8.50 \times 10^7$ & $\bm{0.82 \times 10^7}$ & \revision{41}\\
SAC & 41:25 & $-2.67 \times 10^8$ & $-4.43 \times 10^8$ & $-1.76 \times 10^8$ & --\\
\hline
\end{tabular}
\end{table*}

\begin{figure*}
\subfloat[DDPG-HER\label{fig:FetchDDPG}]{
  \includegraphics[width=0.24\textwidth]{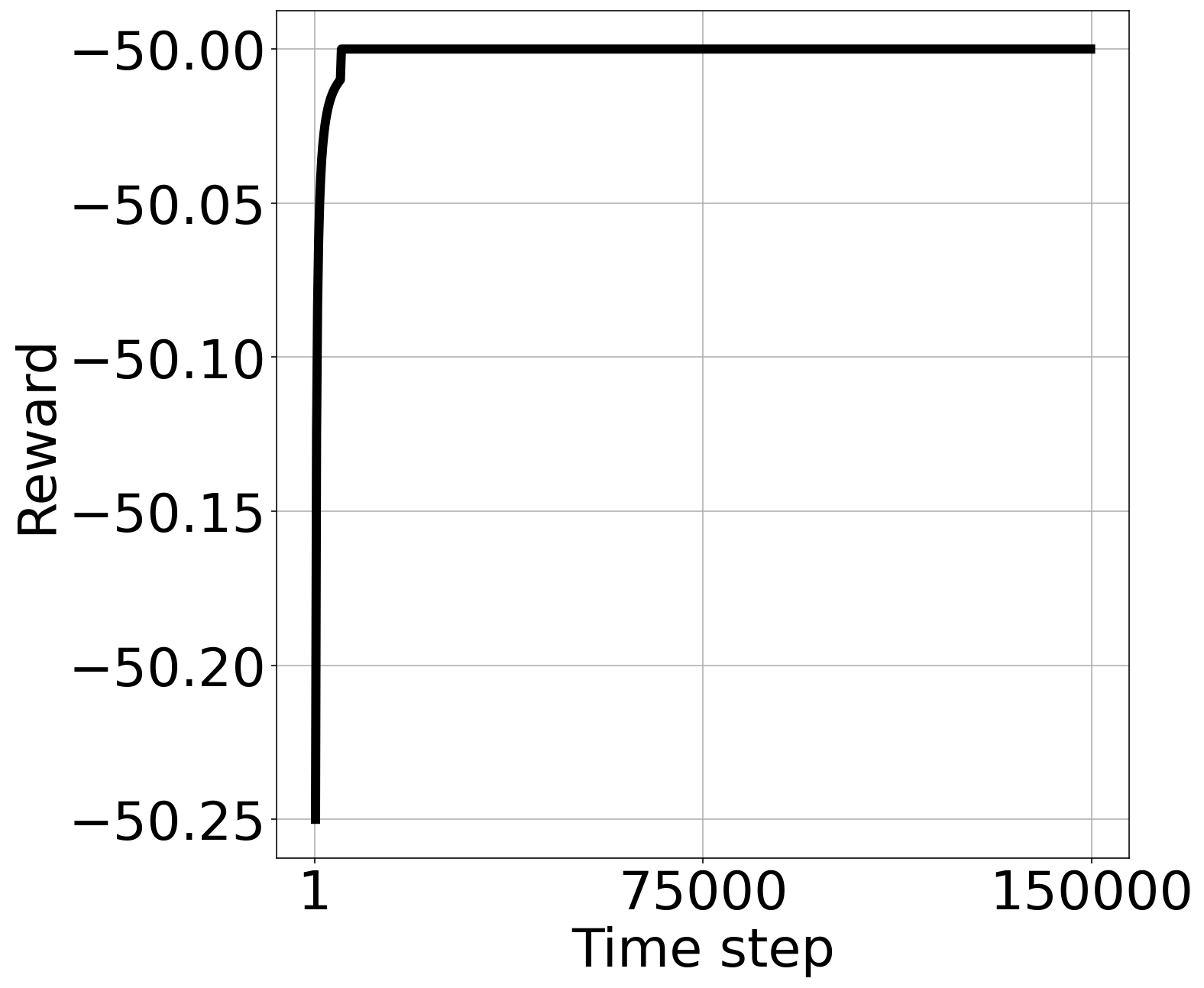}
  }
\hspace{0.1cm} 
\subfloat[SAC-HER\label{fig:FetchSAC}]{
  \includegraphics[width=0.24\textwidth]{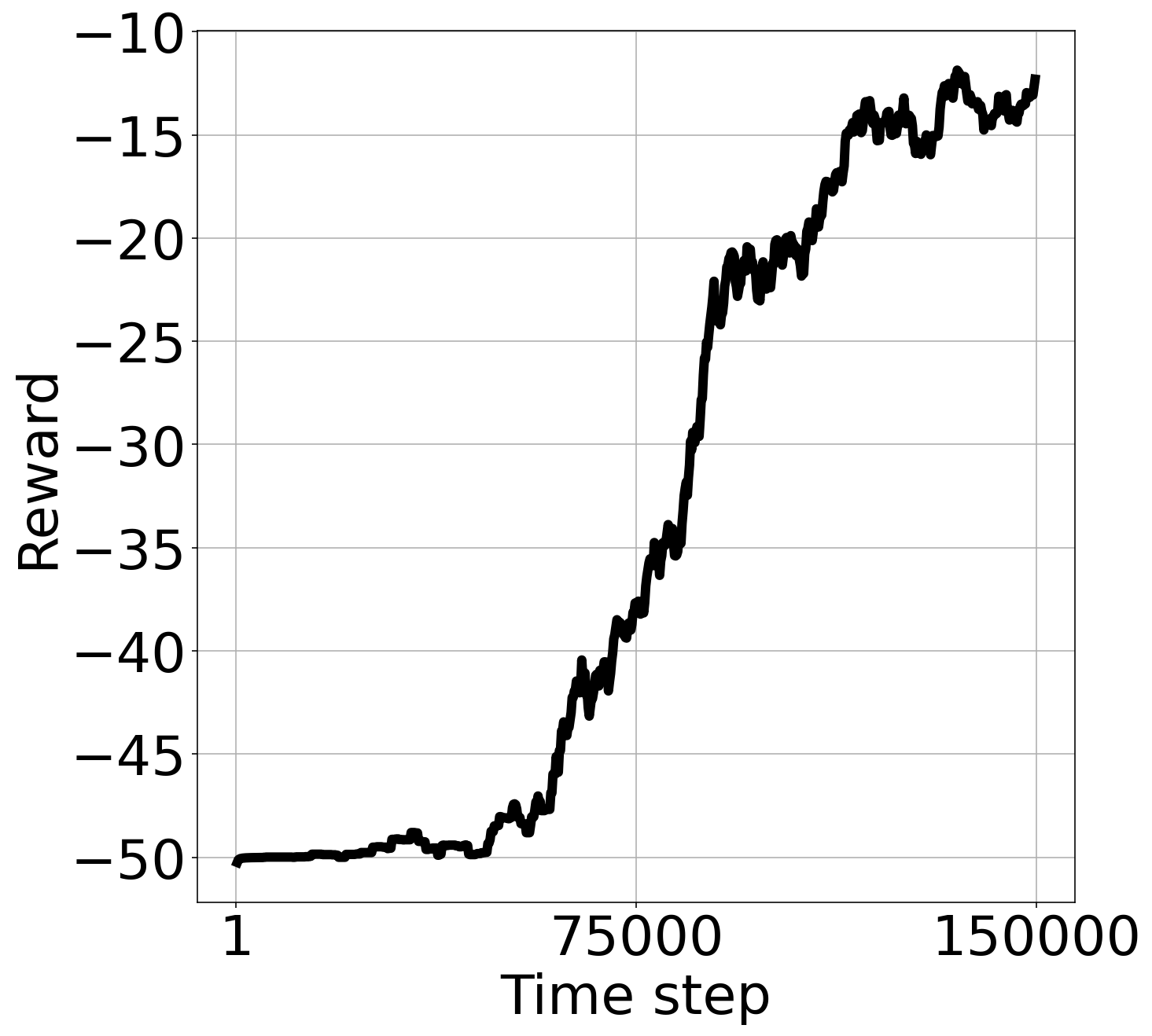}
  }
\subfloat[TD3-HER\label{fig:FetchTD3}]{
  \includegraphics[width=0.24\textwidth]{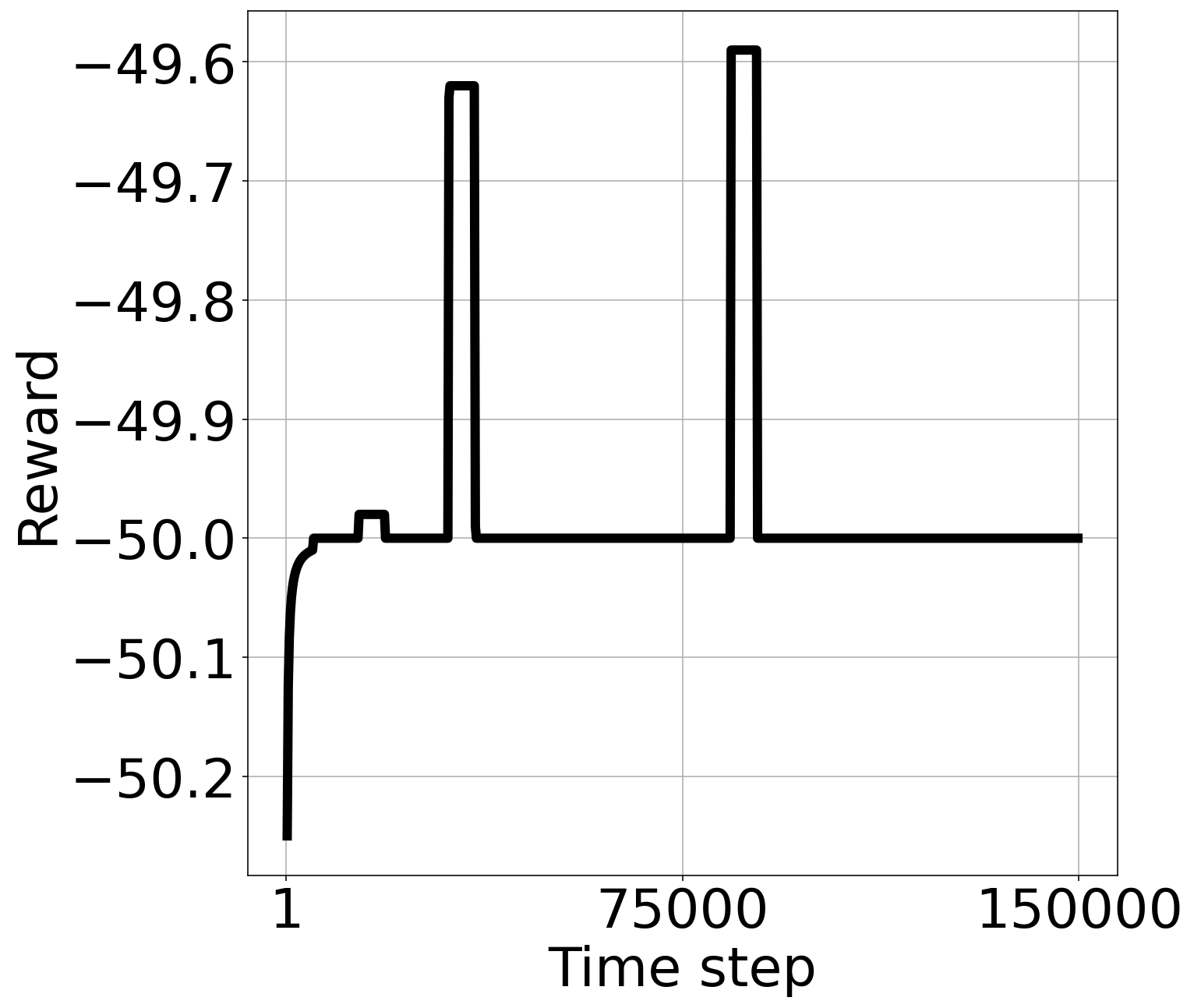}
  }
\subfloat[TQC-HER\label{fig:FetchTQC}]{
  \includegraphics[width=0.24\textwidth]{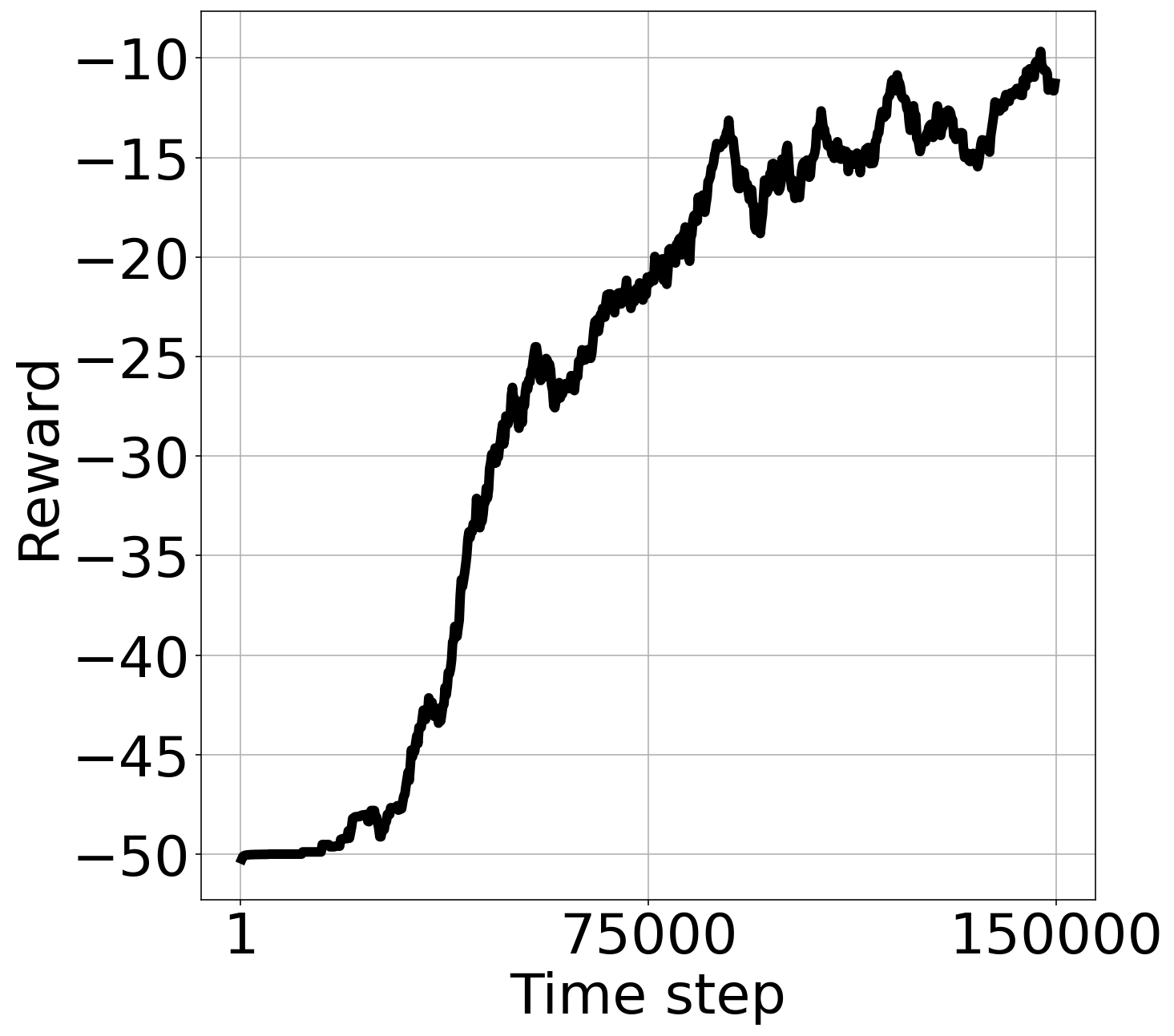}
  }
  \\
\subfloat[A2C\label{fig:incineratorA2C}]{
  \includegraphics[width=0.24\textwidth]{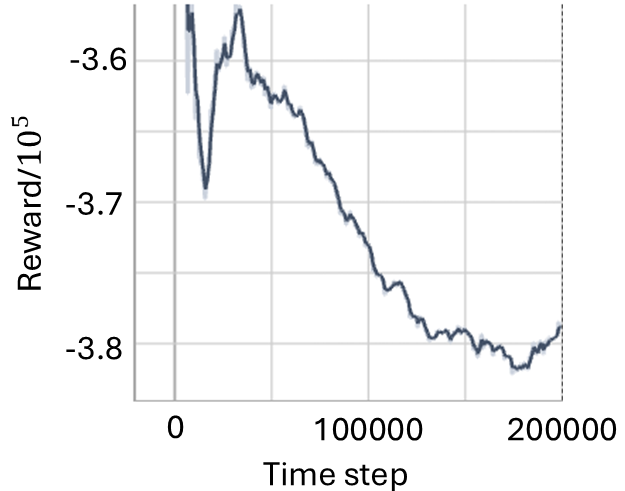}
  }
\subfloat[ARS\label{fig:incineratorARS}]{
  \includegraphics[width=0.24\textwidth]{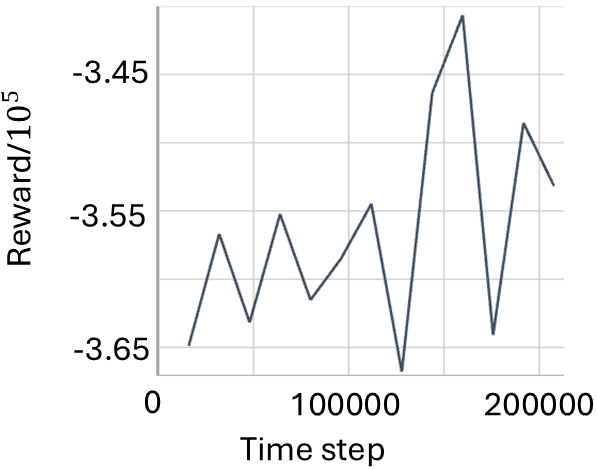}
  }
\subfloat[DDPG\label{fig:incineratorDDPG}]{
  \includegraphics[width=0.24\textwidth]{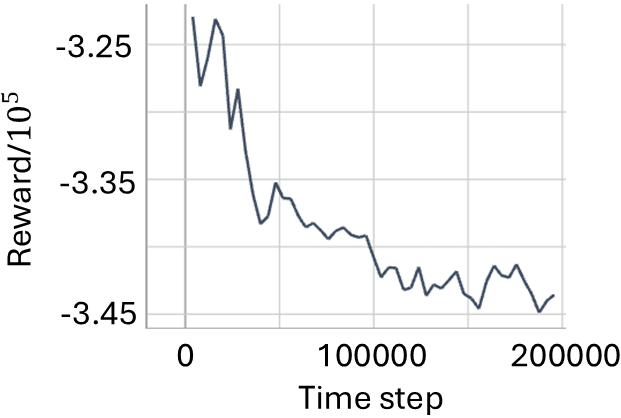}
  }
\subfloat[PPO\label{fig:incineratorPPO}]{
  \includegraphics[width=0.24\textwidth]{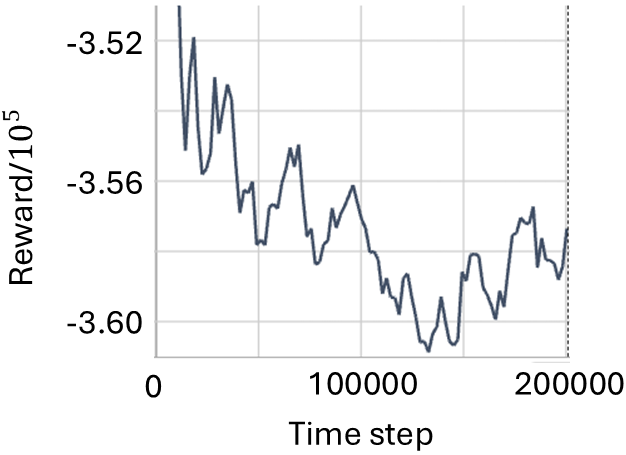}
  }
\\
\subfloat[A2C\label{fig:MonodA2C}]{
  \includegraphics[width=0.24\textwidth]{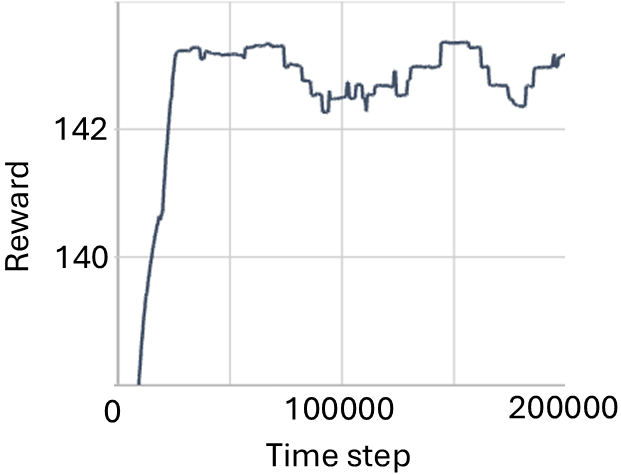}
  }
\subfloat[ARS\label{fig:MonodARS}]{
  \includegraphics[width=0.24\textwidth]{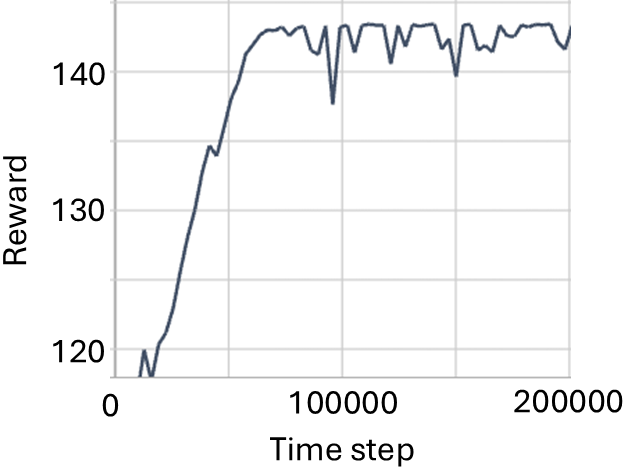}
  }
\subfloat[PPO\label{fig:MonodPPO}]{
  \includegraphics[width=0.24\textwidth]{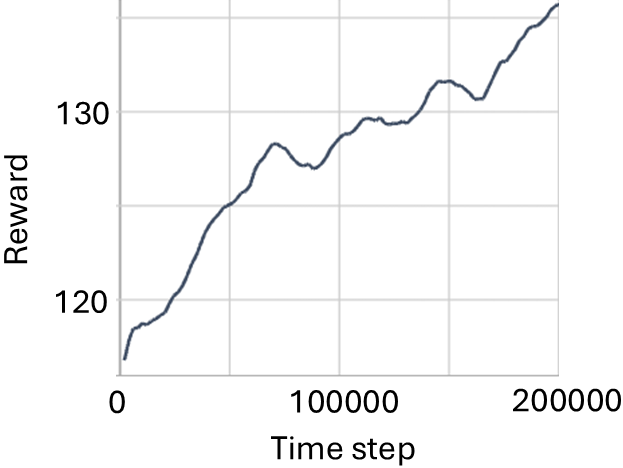}
  }
\subfloat[SAC\label{fig:MonodSAC}]{
  \includegraphics[width=0.24\textwidth]{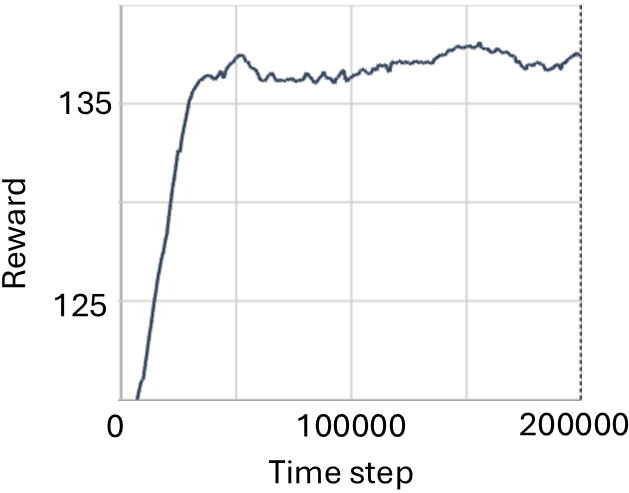}
  }
\\
\subfloat[A2C\label{fig:CarboNetA2C}]{
  \includegraphics[width=0.24\textwidth]{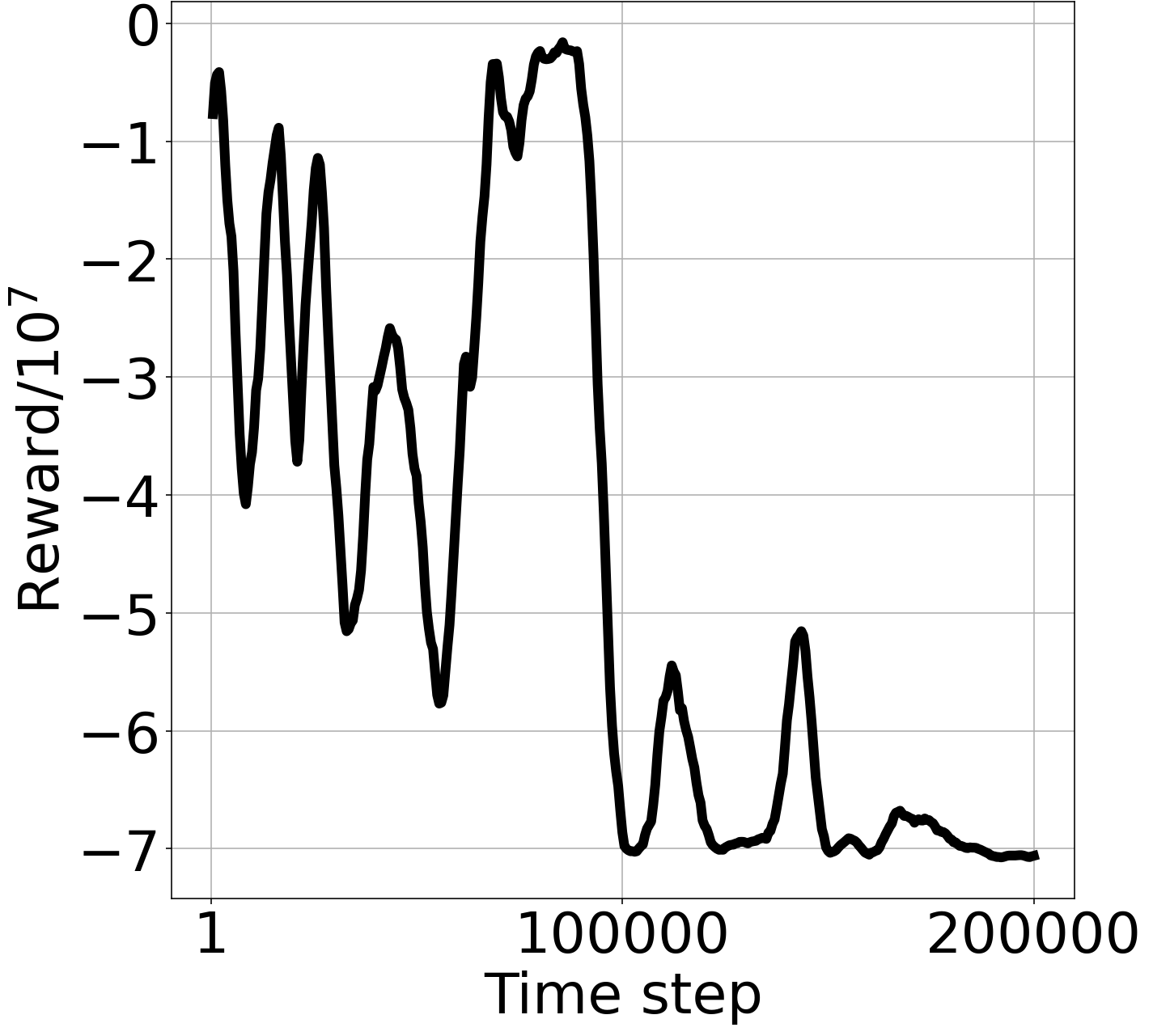}
  }
\hspace{0.1cm} 
\subfloat[ARS\label{fig:CarboNetARS}]{
  \includegraphics[width=0.24\textwidth]{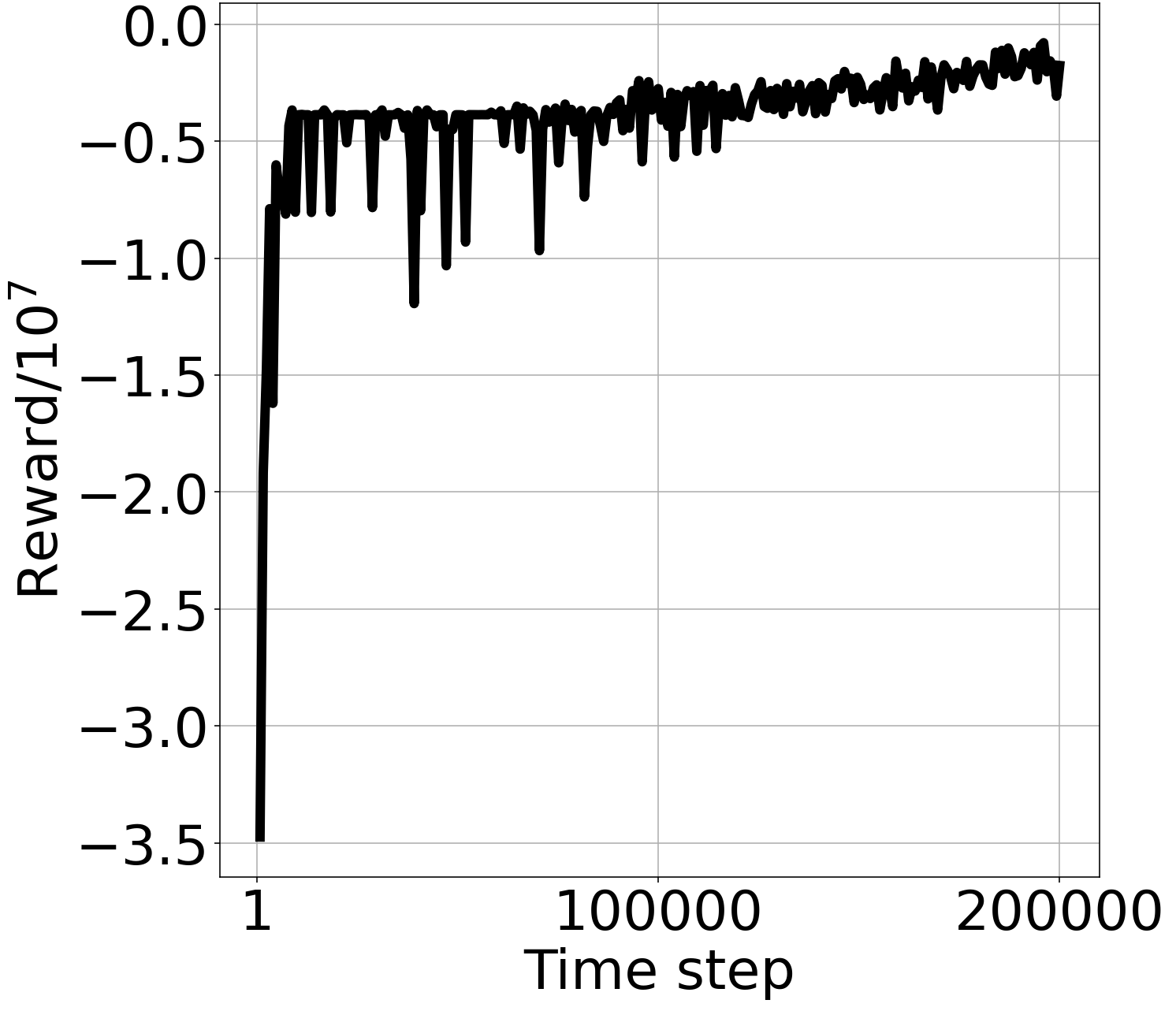}
  }
\subfloat[DDPG\label{fig:CarboNetDDPG}]{
  \includegraphics[width=0.24\textwidth]{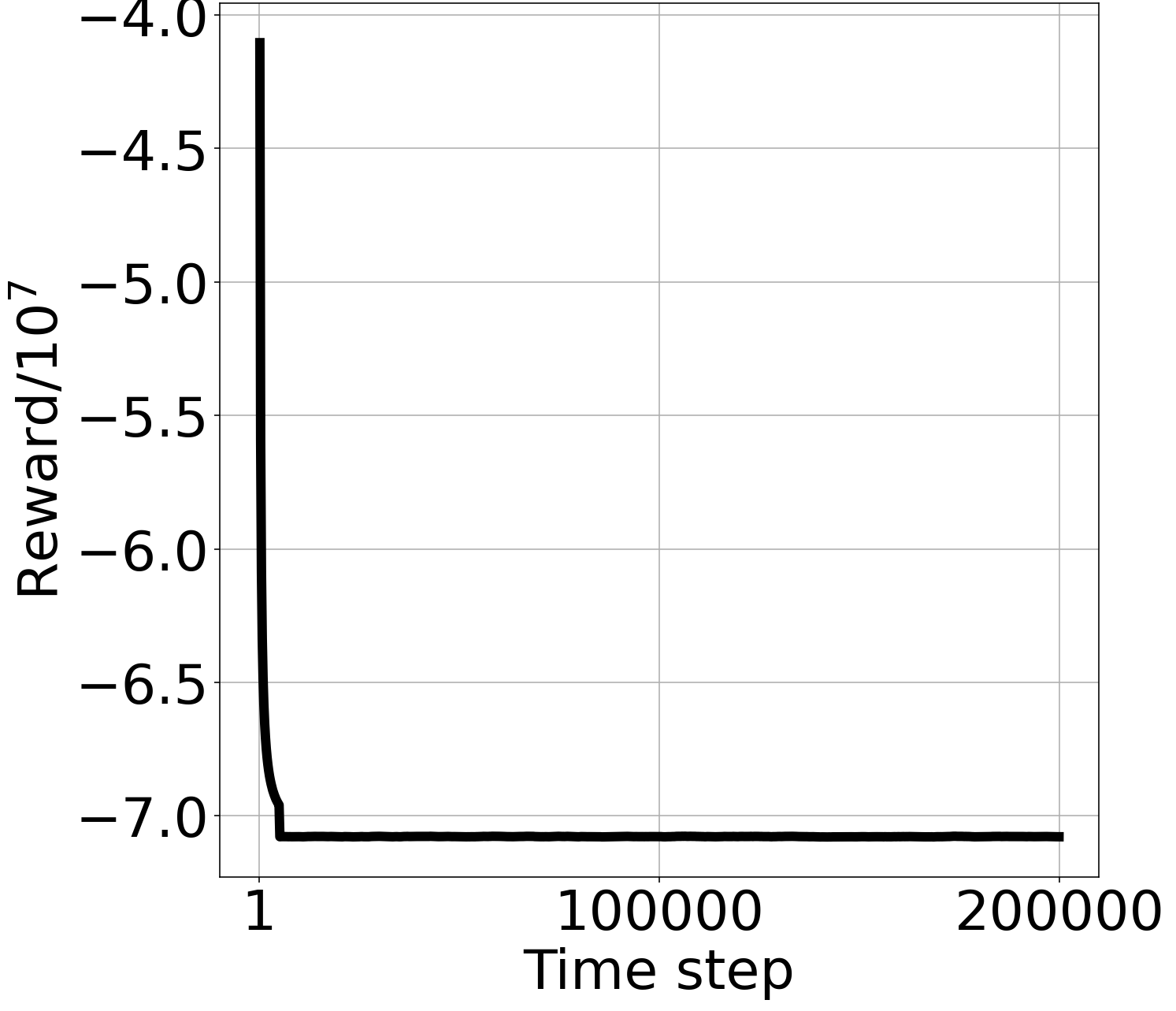}
  }
\subfloat[PPO\label{fig:CarboNetPPO}]{
  \includegraphics[width=0.24\textwidth]{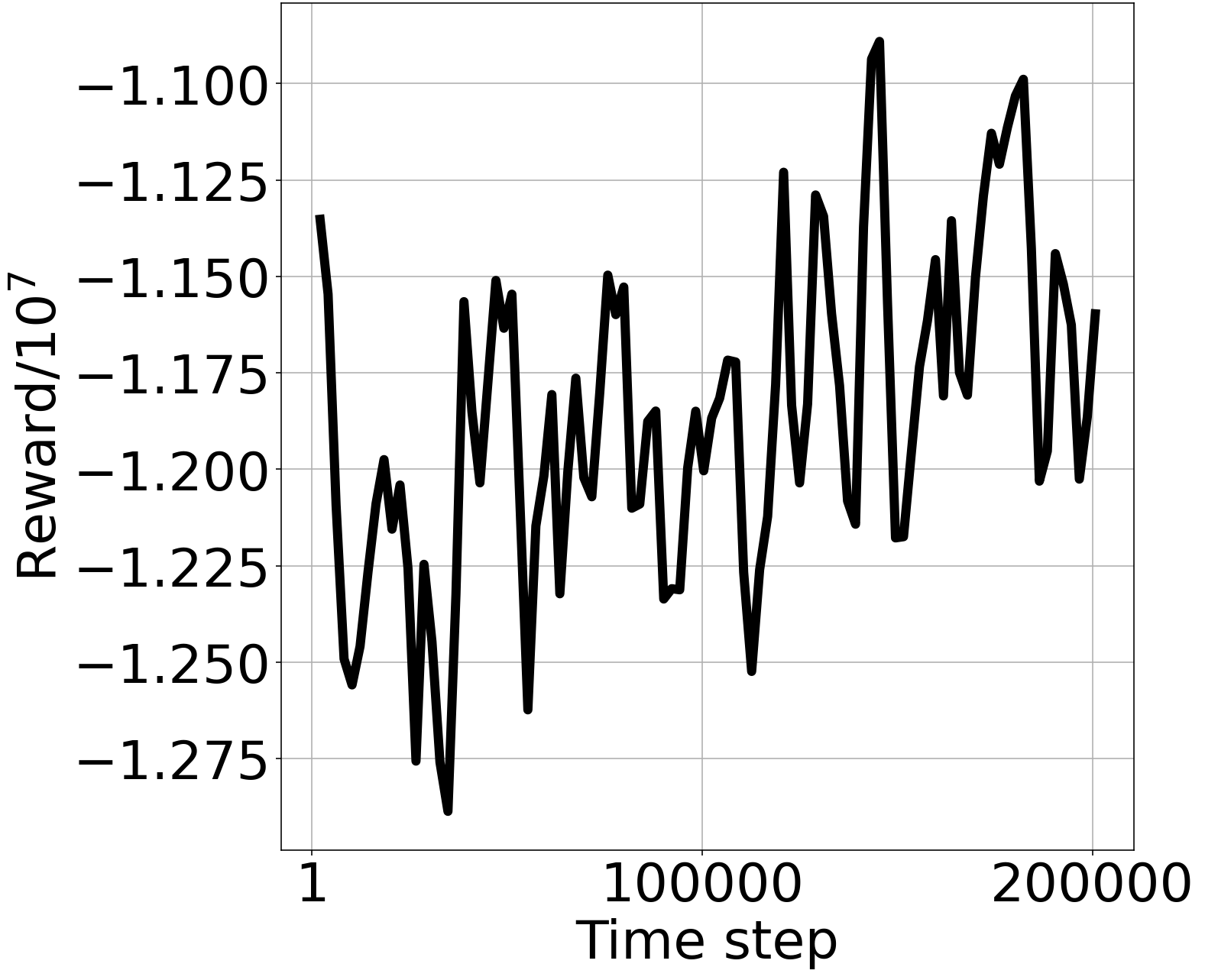}
  }
\\
\subfloat[A2C\label{fig:CopperNetA2C}]{
  \includegraphics[width=0.24\textwidth]{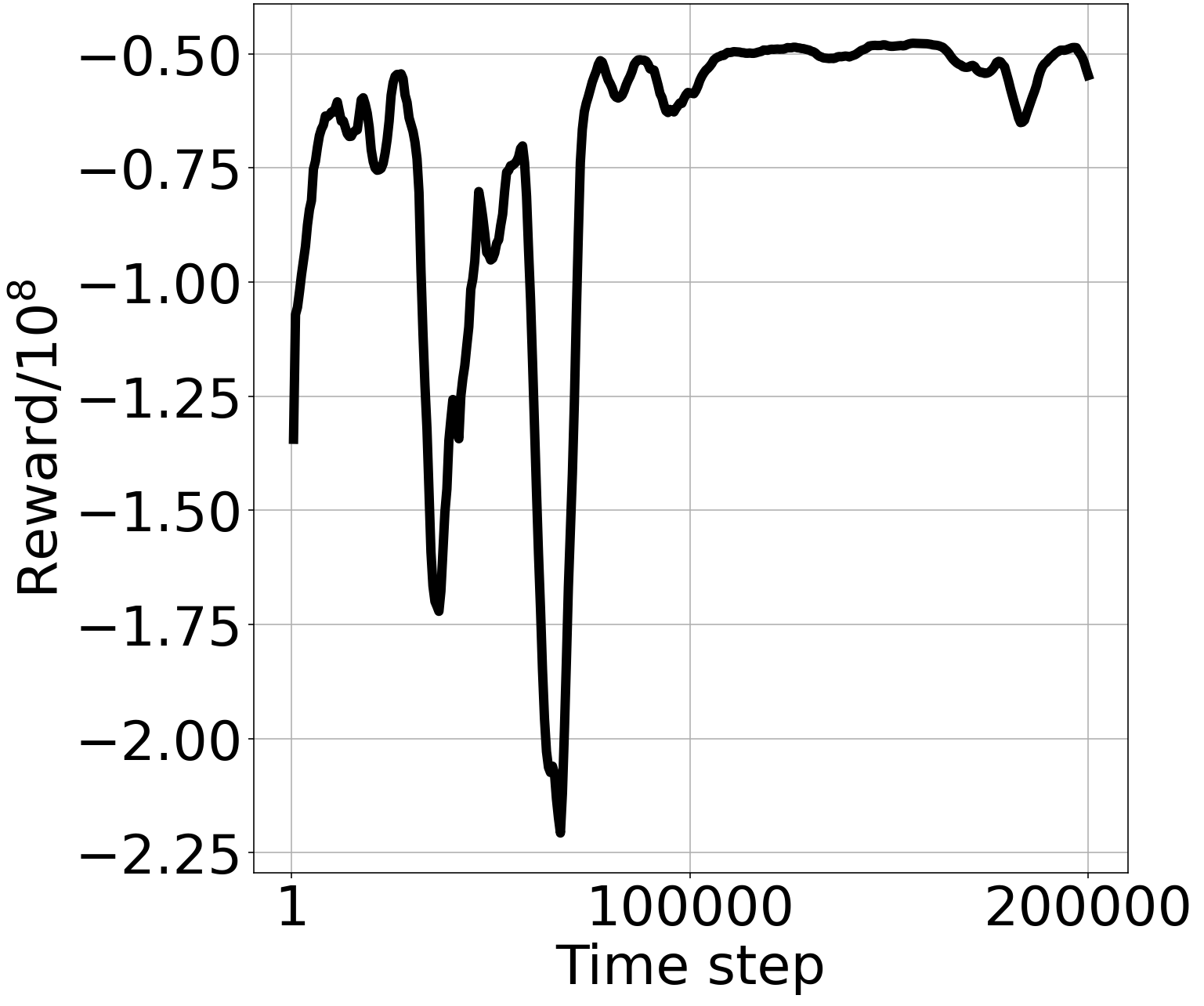}
  }
\subfloat[ARS\label{fig:CopperNetARS}]{
  \includegraphics[width=0.24\textwidth]{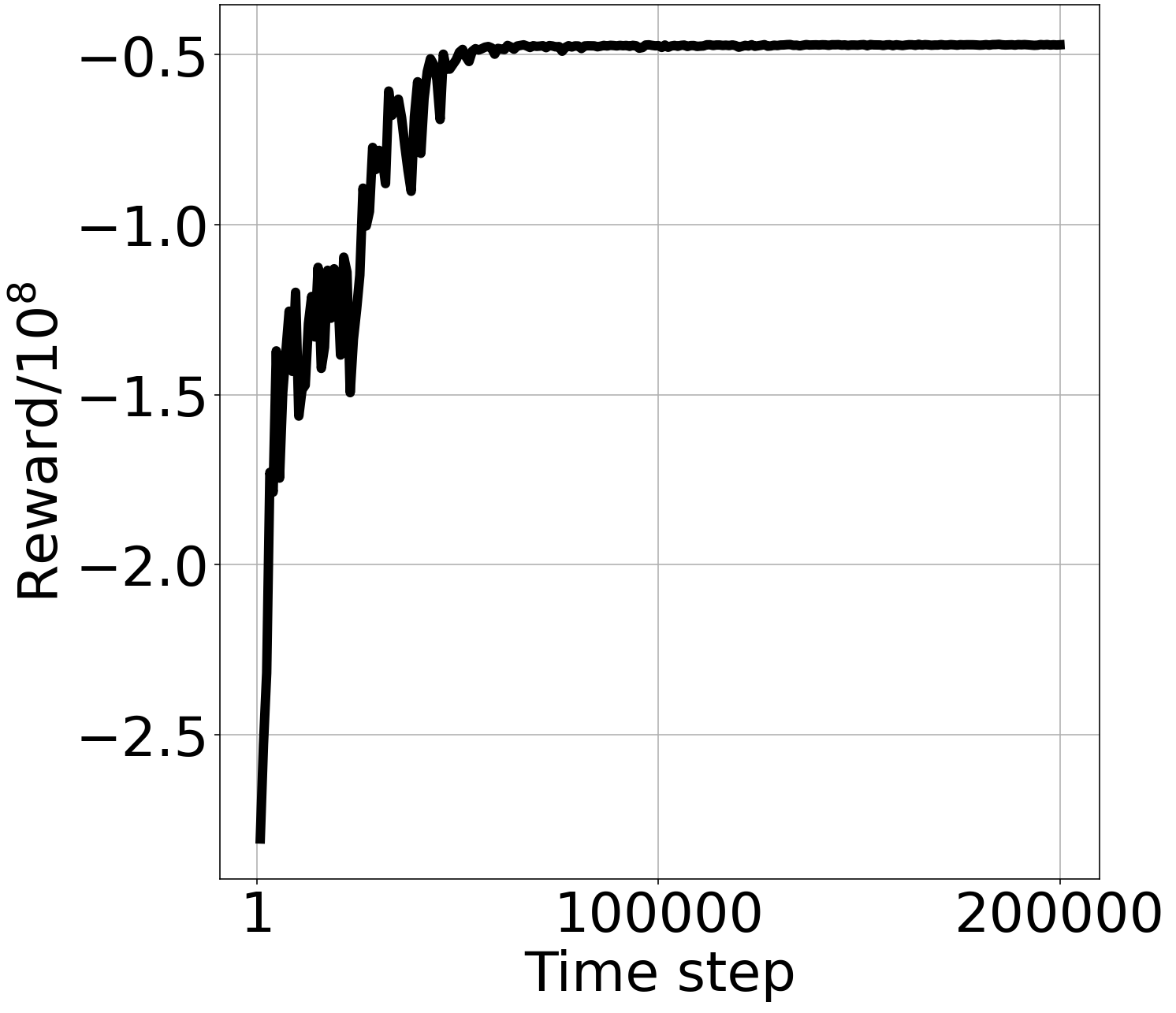}
  }
\subfloat[DDPG\label{fig:CopperNetDDPG}]{
  \includegraphics[width=0.24\textwidth]{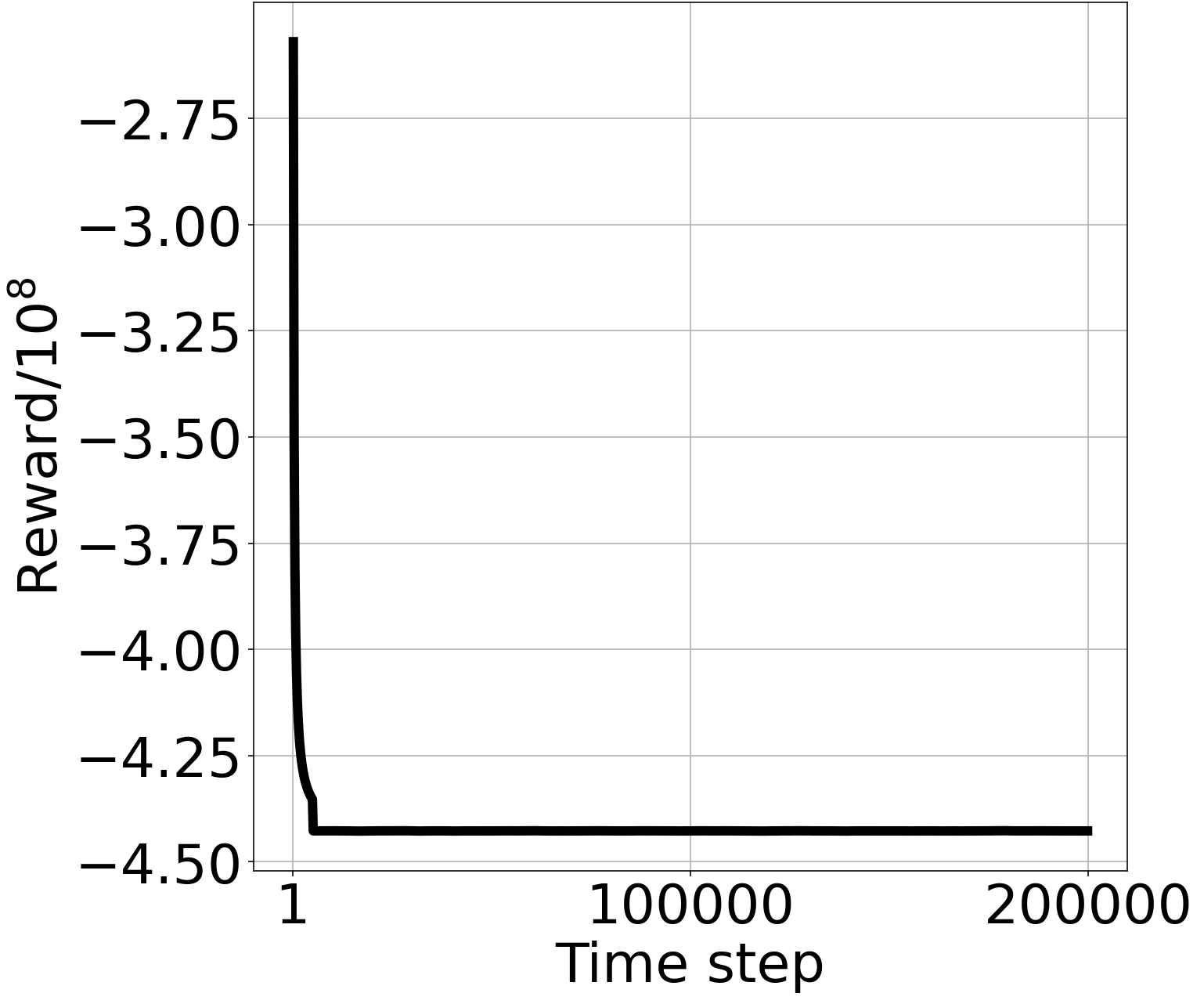}
  }
\subfloat[PPO\label{fig:CopperNetPPO}]{
  \includegraphics[width=0.24\textwidth]{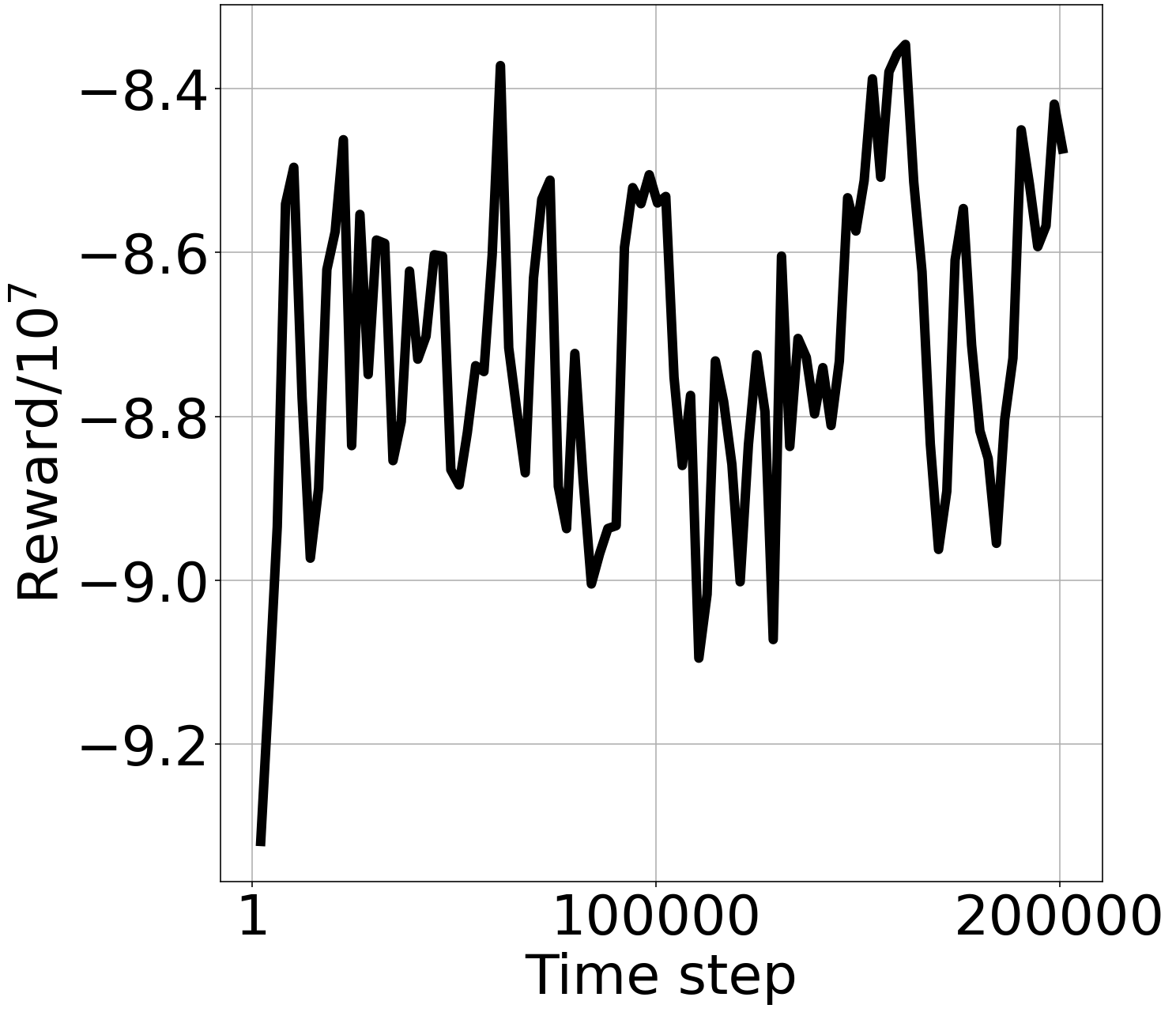}
  }
\caption{\revision{Learning curves of different algorithms with the following environments: (a)-(d) \emph{FetchWasteSorting}; (e)-(h) \emph{Incinerator}; (i)-(l) \emph{CO2MicroalgaeMonod}; (m)-(p) \emph{CarboNet}; (q)-(t) \emph{CopperNet}. The learning curves yield a qualitative evaluation of the convergence stability \citep{rahim2024tinyfdrl}.}}
\label{fig:LearningCurves}
\end{figure*}
}

\section{Suggestions for Users and Developers}\label{sec:forUseAndDev}
\revision{The aim of CiRL is to serve as a tool for decision makers interested in material flow circularity for a more sustainable management of natural resources. Specifically, a trained agent can be implemented in the compartmental network $\mathcal{N}$ and generate a sequence of actions. In practice, the actions produced by the agent are discrete-time values of material flows that can be replicated in real scenario. Clearly, the quality of the actions is as good as the realisticness of the RL environment that the agent has been trained on. At the current stage, our RL environments are provided as paradigmatic examples and do not model any real systems with fidelity; this has been necessary to introduce the concept of CiRL, and hence, to illustrate the use of state-of-the-art RL in the design of circular systems. Making these RL environments more realistic is essential to make CiRL useful for practical scenarios. A guide on how to use and develop CiRL is the scope of this section.}

The general procedure \emph{to use} CiRL is as follows. Note that CiRL does not require to have a GPU nor the drivers and library dependencies configured on your local computer; Colab takes care of them under the hood. 
\begin{enumerate}
\item{Choose the material, namely, $\beta$, whose circularity $\lambda$ (\ref{eq:circularity}) has to be maximized, e.g., a plastic for medical applications, a rare-Earth element, or water.} 
\item{Define the network of thermodynamic compartments $\mathcal{N}_\beta$ processing $\beta$ (Definition \ref{def:TMN}).}
\item{From the compartments available in CiRL, select those that are nodes and arcs of $\mathcal{N}_\beta$.} 
\item{Check that the values of the parameters used in the compartment models are suitable for your case study, e.g., has the truck the mass that you need? Are the initial conditions of the microalgae cultivation tuned as in your case? Are the torque saturation values for the robotic waste sorter correct for your case?}
\item{Define the reward function of each compartment. Since the reward function is a mathematical description of the task you expect to be performed by the compartment, the reward should, directly or indirectly, seek to increase the circularity $\lambda$ (\ref{eq:circularity}) of your network $\mathcal{N}_\beta$ (an example with robotic waste sorting is discussed in Section \ref{sec:Compartments}).} 
\item{Choose an RL algorithm available in SB3 to train a controller optimizing the reward, and hence, directly or indirectly, optimize $\lambda$.} 
\end{enumerate}

\emph{To develop} the library, one should consider to follow the procedure described above for a user with the only difference that, in this case, one or more thermodynamic compartments of $\mathcal{N}_\beta$ are not available in CiRL, and hence, one needs to implement them. In the case one needs a robotic compartment, the best way to proceed is to leverage and modify as needed an existing physics engine as we did in \emph{FetchWasteSorting} and \emph{Reacher} (Table \ref{tab:summaryEnvsAlgs}). In contrast, the literature lacks of RL physics engines for non-robotic systems such as those implemented in CiRL, i.e., the incinerator, the transportation truck, and the microalgae cultivations. In the non-robotic case, the advice is to formulate the compartment model in the state-space form similarly to CiRL since it is a standard format used by the automatic control community useful for systems analysis and controller designs. The ordinary differential equations in the state-space form have to be derived from mass balances and/or the laws of thermodynamics in order to yield a thermodynamic compartment of a TMN (Definition \ref{def:TMN}), and hence, to align with the CiRL framework. This is always possible thanks to the generality of thermodynamics \citep{haddad2017thermodynamics}, although in some cases it can be particularly challenging. Then, the integration step size for numerical integration of the differential equations must be tuned as we did, for example, with \emph{TransportTruck} and \emph{CO2MicroalgaeDroop} (Table \ref{tab:summaryEnvsAlgs}). The suitability of the integration step size can be verified by comparing the state trajectories produced by implementing the model as an SB3 environment (numerically solved via, for example, the Euler's method) and the state trajectories given by the Python solver \emph{scipy.integrate.odeint()} for the same initial conditions and the same control input (i.e., the action). This is what we did, for example, with \emph{TransportTruck} and \emph{CO2MicroalgaeDroop}. Once the state-space form is formulated and solved with a numerically-stable method, the reward function has to be defined in order to, directly or indirectly, increase the circularity $\lambda$ of $\mathcal{N}_\beta$. Finally, one or more SB3 algorithms must be called to train and test the RL controller optimizing the circularity-oriented reward. The newly implemented compartments could be added to CiRL to expand the library for future uses.

\revision{
\section{Applications and Impact}\label{sec:ApplicationsAndImpact}
CiRL is a library developed as a support tool in the design of circular systems. The need of such systems can be considered strategic in the increasing competition between nations to secure access to critical raw materials necessary to make green, digital, and defense technologies \citep{competition1,nygaard2023geopolitical}. The traditional (i.e., linear) economy has also problems at the output stage since it generates large volumes of waste whose management is damaging the natural environment \citep{egger2025evaluating,cottom2024local}. Therefore, actions to increase the circularity of materials is gaining importance \citep{macleod2024waste}, even more considering the fact that the world population is expected to increase and reach approximately 10 billion in 2060 \citep{owid-un-population-2024-revision}. 

The purpose of RL is to generate AI-driven actions so that the RL environment behaves as desired. By developing CiRL, which aims to optimize the systems circularity, we pave the way for generating AI-driven decisions to increase circularity. The state-of-the-art approach to decision making for circularity is material flow analysis (MFA), which is essentially a methodology based on the analysis of large data of material stocks and flows \citep{cullen2022material,brunner2016handbook,EU-MFA,CSIRO-MFA}. Hence, the traditional MFA approach consists of actions for circularity defined by humans formulated on the analysis of data. In constrast, CiRL leverages ordinary differential equations to define the models and, via RL, generates AI-driven actions for circularity. Thus, CiRL is a tool that decision makers can use in combination with MFA, especially considering the pace of improvement that AI systems have been showing in the last fifteen years. 

The CiRL core is a general framework that can be applied to any sector, e.g., aerospace, healthcare, automotive, defense. Each sector involves the supply-recovery chain of several materials and products; thus, targeting a sector of interest requires to focus on the life-cycle stages, i.e., compartments, of the supply-recovery chains of its materials and products, and hence, to develop the CiRL environments modeling the material life-cycles of the target sector.    

The current size of the RL environments in CiRL is limited, and hence, not suitable for realistic decision making. However, CiRL is modular, and hence, it is meant to be scaled-up to target specific real-world scenarios. Given that CiRL is the first library to sit between AI-driven decision making and circular economy, and given that leading European organizations are seeking circular designs solutions based on digital technologies \citep{DiCE-Lab,DICENetwork+}, the expected impact of CiRL is relatively high.
}

\section{Limitations and Future Work}\label{sec:Limitations}
The proposed library, namely, CiRL, paves the way for deep reinforcement learning in a circular economy. This has been possible first of all because we defined a clear measure of circularity, namely, $\lambda(\mathcal{N};t)$, and also because the definition of $\lambda(\mathcal{N};t)$ is predicated on the networked framework in Definition \ref{def:TMN} that enabled us to RL-control one compartment at a time. 

The current limitation of CiRL is that $\lambda(\mathcal{N};t)$ does not enter \emph{explicitly} the reward functions of the environments, which would be the most natural way to optimize circularity via reinforcement learning. In contrast, the reward functions of the existing environments affect $\lambda(\mathcal{N};t)$ \emph{indirectly}. Improving CiRL by including the control of multi-compartment systems optimizing $\lambda(\mathcal{N};t)$ \emph{directly} is one of the main future research directions \revision{along with testing the trained agents in generating actions $u$ when interacting with the environment in a simulation loop. This is intended to be the first version of CiRL.} 

\revision{Other directions of future work are possible: 
\begin{itemize}
\item{Compare the performance of these RL-based controllers with techniques based on control theory, e.g., the linear quadratic regulator (LQR) \citep{zocco2025carbonet}.}
\item{Develop inverse reinforcement learning systems to infer better reward functions for real-world applications \citep{lanzaro2025evaluating,feng2025deep}.}
\item{Develop multi-agent RL agents to control multiple compartments simultaneously; for multi-agent RL, there might be libraries more suitable than SB3, e.g., MARLlib \citep{hu2023marllib}}.
\item{Integrate explainability tools into CiRL to increase the interpretability of the RL agent actions; this is important in the case the AI-generated actions are considered for use in real case studies.}
\item{Create and control compartmental networks targeting rare-Earth elements \citep{Rare-EU} or critical raw materials \citep{CRM-EU} similarly to what we did with CopperNet.}
\end{itemize} 


\section{Conclusion}\label{sec:Conclusion}
In this paper, we proposed CiRL, a library of reinforcement-learning environments that, by leveraging the learning algorithms in Stable-Baselines3, can provide AI-driven actions for optimizing the circularity of supply-recovery chains to be combined with human-driven decisions derived from material flow analysis (MFA) studies. CiRL currently consists of eight environments on which nine RL algorithms were tested. The results demonstrate the usability and scalability of CiRL. The main future work is to enhance the environment fidelity for specific case studies and to design reward functions optimizing circularity explicitly.    
}

\bibliographystyle{elsarticle-harv}
\bibliography{bibliography}

\end{document}